\documentclass[twocolumn]{aastex631}

\usepackage{graphicx}
\usepackage{amsmath}
\usepackage{color}
\usepackage{ulem}
\usepackage{float}
\usepackage{soul}
\usepackage{rotating}
\usepackage[graphicx]{realboxes}
\graphicspath{{fig/}}

\newcommand\lx{$L_{\rm 0.5 - 2\,keV}^{\rm gas}$}

\begin{document}

\title{Diffuse X-Ray-emitting Gas in the Central Region of Star-Forming Galaxies}
\correspondingauthor{Junfeng Wang}
\email{jfwang@xmu.edu.cn}

\author[0009-0007-7542-1140]{Chunyi Zhang}
\affiliation{Department of Astronomy, Xiamen University, 422 Siming South Road, Xiamen 361005, People's Republic of China}

\author[0000-0003-4874-0369]{Junfeng Wang}
\affiliation{Department of Astronomy, Xiamen University, 422 Siming South Road, Xiamen 361005, People's Republic of China}


\begin{abstract}

The interstellar medium of galaxies, with temperatures reaching several million degrees, provides a pivotal perspective for understanding the physical and chemical properties of star formation, galactic evolution, and their associated feedback mechanisms. We use archival data from $Chandra$ observations to extract the diffuse X-ray emission from 23 nearby star-forming galaxies and study its correlation with star formation activity in the central region of these galaxies. The surface brightness profile of each galaxy presents a sharp decrease in the central region of $\thicksim$0.3$-$2 kpc and then varies slowly outside this range. Compared to the global relation between the diffuse thermal X-ray luminosity from hot gas ({\lx}) and the star formation rate (SFR), we found a super-linear relation of ${\rm log}(L_{\rm 0.5-2\,keV}^{\rm gas} /{\rm erg\,s^{-1}})=1.34\,{\rm log}({\rm SFR}/{M_{\rm \odot}}\,{\rm yr^{-1}})+40.15$ for the center of these sample galaxies. This result suggests that more intense stellar feedback is associated with stronger star formation activity in the central region of star-forming galaxies, where more energy output from supernovae (SNe) and stellar winds is converted into X-ray flux. Furthermore, the slope of the {\lx}$-$SFR relation anticorrelates with spatial scale in the galactic central region. This indicates that the characteristics of central hot gas emission are gradually averaged over larger areas. The diffuse X-ray luminosity also shows a good correlation with molecular gas, stellar mass, and mid-plane pressure traced by the baryonic mass, although these relations show relatively large scatter.

\end{abstract}


\section{Introduction}\label{sec:intro}

The diffuse X-ray emission of hot interstellar medium is an essential tool for studying the feedback of star formation and galactic evolution. Early X-ray images of late-type galaxies have shown that the spatial distribution of diffuse X-ray emission correlates with the sites of recent star formation \citep{2002_Fraternali_ApJ_109F,2004_Tyler_ApJ,2004_Doane_AJ_2712D,2004_Strickland_ApJS_193S}. Deep $Chandra$ observations of M101 also found that over 80\% of the diffuse emission is associated with star-forming regions younger than 20 Myr as traced by far-ultraviolet (UV) emission \citep{2010_Kuntz_ApJS_46K}. At the global scale, recent studies revealed a linear correlation between the total X-ray luminosity and the star formation rate of the host galaxy \citep{2003_Ranalli_A&A,2005_Grimes_ApJ_187G,2009_Owen_MNRAS_1741O,2012_Mineo_hotgas,2012_Mineo_HMXBs,2013_LiJiangTao_MNRAS_2,2014_Mineo_MNRAS_1698M,2018_Smith_AJ_81S}. These well-established relations imply that the origin of the X-ray emission must be closely related to star formation.

Nearby star-forming galaxies provide us with the opportunity to investigate the properties of the hot ionized interstellar gas and its origin on smaller spatial scales. In contrast to the linear correlation between X-ray emission and SFR at the global scale, the $L_{\rm X}$$-$SFR relation shows varying slopes and intercepts across different regions of galaxies on (sub)kiloparsec scales \citep{2010_Kuntz_ApJS_46K,2020_Kouroumpatzakis_mnras,2024_Zhangcy_ApJL,2025_Zhangcy_ApJ_15Z}. The study of individual star-forming regions in the galactic disk, giant H\,$_{\rm \uppercase\expandafter{\romannumeral2}}$ regions, and the bulge of M101 found that the diffuse X-ray surface brightness is roughly proportional to the square root of the FUV surface brightness \citep[e.g.,][]{2010_Kuntz_ApJS_46K}. Furthermore, the FUV/X-ray correlation in M101 varies with galactic radius and shows a large scatter in different radial ranges. This suggests a significant influence of local environment for the thermal X-ray brightness and star formation activity. 

A similar study on diffuse X-ray emission has been carried out in the nuclear regions of five late-type spiral galaxy samples \citep[e.g,][]{2024_Zhangcy_ApJL}. The high spatial resolution of $Chandra$ allows us to measure X-ray flux on the scale of a few hundred parsecs for these five nearest galaxies. Contrary to expectations, \citet{2024_Zhangcy_ApJL} obtain a super-linear relation between SFR and diffuse X-ray luminosity in the 0.5$-$2 keV band, with a power-law index of $n \approx 1.8$. Again, \citet{2004_Tyler_ApJ} found a comparable trend between diffuse X-ray and mid-infrared (IR) fluxes from 12 galaxy cores, with X-ray flux spanning three orders of magnitude and mid-IR flux spanning two orders of magnitude. Compared to the spiral arms, this result shows a larger scatter in X-ray and mid-IR fluxes, implying that more than one factor contributes to the central diffuse X-ray emission. These could include the depth of the gravitational potential in the bulge, stellar winds and supernova feedback originating from bulge populations, active galactic nucleus (AGN) activity, and historical star formation episodes \citep{1989_Fabbiano_ARA&A_87F,2004_Tyler_ApJ,2010_Hopkins_MNRAS_7H,2019_Fabbiano_cxro.book_7F,2022_Nardini_hxga}. A systematic high spatial resolution survey of the X-ray emission remains essential for determining the physical origin and intrinsic properties of the diffuse hot gas. 

Here we present a detailed analysis of the relationship between central diffuse X-ray emission and SFR for 23 star-forming galaxies. All these galaxies are selected from the Physics at High Angular Resolution in Nearby Galaxies (PHANGS) survey \citep{2021_Leroy_ApJS}. This survey provides sensitive, high-resolution, wide-field imaging data across multiple wavelengths to quantify the physics of star formation and feedback at the scale of molecular clouds. Thus, these comprehensive data are well suited to help study the diffuse X-ray emission and its relationship to stellar feedback.

In this paper, we aim to investigate the scaling relation of {\lx} with SFR on different spatial scales and examine its scale dependence in the central region of the 23 sample galaxies. We measured the slope and scatter of the resolved {\lx}$-$SFR relation using a Bayesian method. We study the effects of molecular gas and stellar mass on the diffuse thermal emission in the central regions of our sample galaxies. Additionally we estimate the unresolved emission of the faint compact sources and their contribution to the soft X-ray luminosity in the 0.5$-$2 keV band. 

The structure of this paper is as follows: In Section~\ref{sec:data} we describe our sample and the methods used in our data reduction and analysis. In Section~\ref{sec:results} we present our main results. In Section~\ref{sec:discussion} we discuss our findings along with other influencing factors. Finally, our conclusions are summarized in Section~\ref{sec:summary}.

\begin{deluxetable*}{lcccccccccc}
\centering
\tablecaption{Galaxy Sample \label{tab:table1}}
\addtolength{\tabcolsep}{2.0pt}
\tablewidth{0pt}
\tablehead{
\colhead{Galaxy} & \colhead{$d$} & \colhead{$i$} & \colhead{$R_e$} & \colhead{$r_c$} & \colhead{log\,$M_{\rm \star}$} & \colhead{log\,$M_{\rm mol}$} & \colhead{log\,${\rm SFR}$} & \colhead{log\,\lx} & \colhead{$\chi^2$/$\nu$} \\
\colhead{} & \colhead{(Mpc)} & \colhead{(deg)} & \colhead{(kpc)} & \colhead{(kpc)} & \colhead{($M_{\odot}$)} & \colhead{($M_{\odot}$)} & \colhead{($M_{\odot}$ ${\rm yr}^{-1}$)} & \colhead{(erg s$^{\rm -1}$)} & \colhead{} \\
\colhead{(1)} & \colhead{(2)} & \colhead{(3)} & \colhead{(4)} & \colhead{(5)} & \colhead{(6)} & \colhead{(7)} & \colhead{(8)} & \colhead{(9)} & \colhead{(10)} 
}
\startdata
NGC 253 & 3.70 & 75.0 & 4.7 & 1.01 & 9.86 $\pm$ 0.02 & 9.46 $\pm$ 0.01 & 0.42 $\pm$ 0.02 & 40.93 $\pm$ 0.09 & 187.98/90 \\
NGC 628 & 9.84 & 8.9 & 3.9 & 0.28 & 8.75 $\pm$ 0.03 & 7.60 $\pm$ 0.06 & $-$1.78 $\pm$ 0.03 & 37.06 $\pm$ 0.70 & 12.83/10 \\
NGC 1313 & 4.32 & 34.8 & 2.5 & 1.81 & 9.02 $\pm$ 0.02 & ... & $-$0.64 $\pm$ 0.02 & 39.59 $\pm$ 0.26 & 51.87/48 \\
NGC 1433 & 18.94 & 28.6 & 4.3 & 1.60 & 9.82 $\pm$ 0.02 & 9.02 $\pm$ 0.02 & $-$0.82 $\pm$ 0.02 & 39.00 $\pm$ 0.34 & 7.61/5 \\
NGC 1511 & 15.28 & 72.7 & 2.4 & 0.80 & 9.12 $\pm$ 0.03 & 8.22 $\pm$ 0.03 & $-$0.35 $\pm$ 0.03 & 39.22 $\pm$ 0.27 & 4.58/5 \\
NGC 1559 & 19.44 & 65.4 & 3.9 & 0.73 & 9.40 $\pm$ 0.02 & 8.10 $\pm$ 0.04 & $-$0.88 $\pm$ 0.02 & 39.46 $\pm$ 0.26 & 11.04/9 \\
NGC 1637 & 11.70 & 31.1 & 2.8 & 0.99 & 9.12 $\pm$ 0.02 & 8.28 $\pm$ 0.02 & $-$0.41 $\pm$ 0.02 & 39.16 $\pm$ 0.71 & 60.96/73 \\
NGC 1792 & 16.20 & 65.1 & 4.1 & 0.58 & 9.07 $\pm$ 0.02 & 8.69 $\pm$ 0.01 & $-$1.02 $\pm$ 0.03 & 39.19 $\pm$ 0.52 & 10.03/7 \\
NGC 2903 & 9.61 & 66.8 & 3.7 & 0.65 & 9.40 $\pm$ 0.02 & 8.80 $\pm$ 0.01 & $-$0.21 $\pm$ 0.02 & 40.01 $\pm$ 0.16 & 57.62/39 \\
NGC 2997 & 14.06 & 33.0 & 6.1 & 0.91 & 9.23 $\pm$ 0.02 & 8.99 $\pm$ 0.01 & $-$0.49 $\pm$ 0.02 & 38.77 $\pm$ 0.42 & 5.56/4 \\
NGC 3351 & 9.96 & 45.1 & 3.0 & 0.96 & 9.42 $\pm$ 0.02 & 8.68 $\pm$ 0.03 & $-$0.28 $\pm$ 0.02 & 39.47 $\pm$ 0.12 & 35.11/35 \\
NGC 3507 & 23.55 & 21.7 & 3.7 & 1.19 & 9.20 $\pm$ 0.02 & 8.42 $\pm$ 0.04 & $-$1.05 $\pm$ 0.02 & 38.41 $\pm$ 0.39 & 6.95/6 \\
NGC 3521 & 13.24 & 68.8 & 3.9 & 1.01 & 9.89 $\pm$ 0.02 & 8.58 $\pm$ 0.02 & $-$0.71 $\pm$ 0.02 & 39.44 $\pm$ 0.36 & 21.23/15 \\
NGC 4254 & 13.10 & 34.4 & 2.4 & 1.83 & 9.86 $\pm$ 0.02 & 9.20 $\pm$ 0.02 & $-$0.19 $\pm$ 0.02 & 40.16 $\pm$ 0.13 & 18.83/19 \\
NGC 4321 & 15.21 & 38.5 & 5.5 & 1.54 & 9.67 $\pm$ 0.02 & 9.30 $\pm$ 0.02 & $-$0.13 $\pm$ 0.02 & 40.11 $\pm$ 0.20 & 65.71/59 \\
NGC 4457 & 15.10 & 17.4 & 1.5 & 1.71 & 9.80 $\pm$ 0.02 & 8.94 $\pm$ 0.03 & $-$0.61 $\pm$ 0.02 & 39.08 $\pm$ 0.13 & 23.82/31 \\
NGC 4459 & 15.85 & 47.0 & 2.1 & 2.24 & 10.06 $\pm$ 0.02 & 8.19 $\pm$ 0.01 & $-$1.01 $\pm$ 0.02 & 39.50 $\pm$ 0.21 & 13.08/17 \\
NGC 4536 & 16.25 & 66.0 & 4.4 & 1.40 & 9.85 $\pm$ 0.02 & 9.27 $\pm$ 0.01 & 0.33 $\pm$ 0.02 & 40.64 $\pm$ 0.46 & 13.03/8 \\
NGC 4945 & 3.47 & 90.0 & 4.5 & 0.52 & 9.45 $\pm$ 0.02 & 9.35 $\pm$ 0.01 & $-$0.25 $\pm$ 0.02 & 39.99 $\pm$ 0.23 & 108.98/90 \\
NGC 5236 & 4.89 & 24.0 & 3.5 & 0.64 & 9.41 $\pm$ 0.02 & 8.94 $\pm$ 0.01 & 0.12 $\pm$ 0.02 & 40.22 $\pm$ 0.08 & 172.04/90 \\
NGC 5248 & 14.87 & 47.4 & 3.2 & 0.91 & 9.18 $\pm$ 0.02 & 9.04 $\pm$ 0.01 & $-$0.46 $\pm$ 0.02 & 38.92 $\pm$ 0.21 & 10.32/5 \\
NGC 6744 & 9.39 & 52.7 & 7.0 & 0.43 & 9.27 $\pm$ 0.02 & ... & $-$2.13 $\pm$ 0.02 & 37.89 $\pm$ 0.35 & 19.50/20 \\
NGC 7793 & 3.62 & 50.0 & 1.9 & 0.64 & 8.53 $\pm$ 0.03 & 7.16 $\pm$ 0.02 & $-$1.64 $\pm$ 0.03 & 37.60 $\pm$ 0.05 & 37.86/33 \\
\enddata
\tablecomments{ Column (2): distance; column (3): inclination angle; column (4): half-mass radius. Values in columns (2)$-$(4) are from \citet{2021_Leroy_ApJS}. column (5): break radius (see Section~\ref{Radial_profile}); column (6): stellar mass; column (7): molecular gas mass; column (8): star formation rate; column (9): diffuse X-ray luminosity in the 0.5$-$2 keV band. Columns (6)$-$(9) are the properties of the centers of the sample galaxies. column (10): $\chi^2$ divided by the number of degrees of freedom, $\nu$, for the best-fit spectral model.  }
\vspace{-0.3cm}
\end{deluxetable*}

\begin{deluxetable}{lcc}
\tablecaption{X-Ray Data \label{tab:addition}}
\addtolength{\tabcolsep}{1.5pt}
\tablewidth{0pt}
\tablehead{
\colhead{Galaxy} & \colhead{Exp. Time} & \colhead{Data Set(s)} \\
\colhead{} & \colhead{(ks)} & \colhead{}  
}
\startdata
NGC 253 & 449.4 & 3931, 1383(0-2), 23(492-501) \\
NGC 628 & 210.1 & 205(7-8), 1600(2-3) \\
NGC 1313 & 94.2 & 2950, 355(0-1), 14676, 15594 \\
NGC 1433 & 48.6 & 16345 \\
NGC 1511 & 38.8 & 15804 \\
NGC 1559 & 45.4 & 16745  \\
NGC 1637 & 211.7 & 76(3-6), 19(68-70)  \\
NGC 1792 & 19.8 & 19524  \\
NGC 2903 & 93.6 & 11260  \\
NGC 2997 & 45.5 & 15383  \\
NGC 3351 & 118.5 & 59(29-31)  \\
NGC 3507 & 39.3 & 3149  \\
NGC 3521 & 71.5 & 9552  \\
NGC 4254 & 44.5 & 17462  \\
NGC 4321 & 146.7 & 6727, 9121, 12696, 14230 \\
NGC 4457 & 38.9 & 3150  \\
NGC 4459 & 39.6 & 2927, 11784  \\
NGC 4536 & 14.9 & 19387  \\
NGC 4945 & 378.7 & 864, 13791, 1498(4-5) \\
NGC 5236 & 820.0 & 2064, 1299(2-6), 13202, 13241, \\
 & & 13248, 14332, 14342, 16024 \\
NGC 5248 & 49.1 & 15386  \\
NGC 6744 & 102.3 & 15384, 18454  \\
NGC 7793 & 229.9 & 3954, 14231  \\
\enddata
\tablecomments{ The data are deflared, and the exposure times are good time intervals. }
\vspace{-0.6cm}
\end{deluxetable}

\section{data} \label{sec:data}

We use a sample of 23 star-forming galaxies (non-AGN) from the PHANGS survey \citep{2021_Leroy_ApJS}, all of which have the $Chandra$ observations. The PHANGS survey selects galaxies with distances less than 24 Mpc to resolve the typical scale of star-forming regions and prefers inclinations to be relatively face on to limit the effects of extinction. These galaxies are a representative set of galaxies where most of the star formation in the local Universe is occurring. Our sample is summarized in Table~\ref{tab:table1} where we use the global parameters (distance, inclination, and half-mass radius) as reported by \citet{2021_Leroy_ApJS}.

\subsection{Chandra }\label{sec:chandra}

To study the resolved X-ray emission from the hot ionized interstellar gas, we selected only galaxies with total $Chandra$ exposure times longer than 15 ks from the PHANGS survey. In addition, we also required that the galaxies have at least 150 net counts in the ACIS broad energy band after removing point sources to obtain reliable X-ray spectra and meaningful constraints. Finally, the constructed sample includes 23 nearby galaxies (see Table~\ref{tab:table1}) with distances spanning a range from $\thicksim$3 to $\thicksim$24 Mpc.

All X-ray data used in this work were downloaded from the $Chandra$ Data Archive and reprocessed using CIAO v.4.13 and CALDB v.4.9.6. The good time intervals were identified by the CIAO routine $deflare$ with the sigma clipping command set to nsigma = 3, restricting the energy range to 0.3$-$8 keV. The final $Chandra$ exposure times range from 14.9 ks to 820 ks, with a median time of 71.5 ks. Table~\ref{tab:addition} presents the detailed exposure times for each sample galaxy, along with the data sets used.

\emph{Point Source Detection and Removal.} We combined multiple observations when available for these objects to create merged maps by the CIAO routine \emph{merge$\_$obs}. Using the merged maps or event maps when only a single exposure could be obtained for a galaxy, we performed point source detection with the wavelet-based source detection algorithm \emph{wavdetect} \citep{2002_Freeman_ApJS}. For each galaxy, the detections were performed on the scales of 1, $\sqrt{2}$, 2, 2$\sqrt{2}$, 4, 4$\sqrt{2}$, and 8 pixels ($0^{\prime \prime}.492$/pixel) and in the soft (0.5 ${-}$ 2.0 keV), hard (2.0 ${-}$ 8.0 keV), and total (0.5 ${-}$ 8.0 keV) energy bands. 

Given that the point spread function (PSF) for a single source can vary significantly between observations, point source removal was carried out on an observation-by-observation basis. For most sources, we used the information about the PSF shape at the source position to determine the radius of the region enclosing 90\% of the source counts. For brighter X-ray sources, we manually increased the size of the region to remove visible ring-like features caused by the PSF wings, thereby minimizing contamination from the PSF spillover of these sources. However, the contamination of unresolved faint X-ray sources is difficult to detect and remove from the diffuse emission of hot gas due to the sensitivity limit of the instruments. Although this cannot currently be excluded, we estimate the contribution of faint compact sources to be less than 30\% of the total X-ray luminosity in the 0.5$-$2 keV band (see discussion in Section~\ref{sec:Contamination}). 

\begin{figure*}[htbp]
\centering 
\includegraphics[scale=0.87]{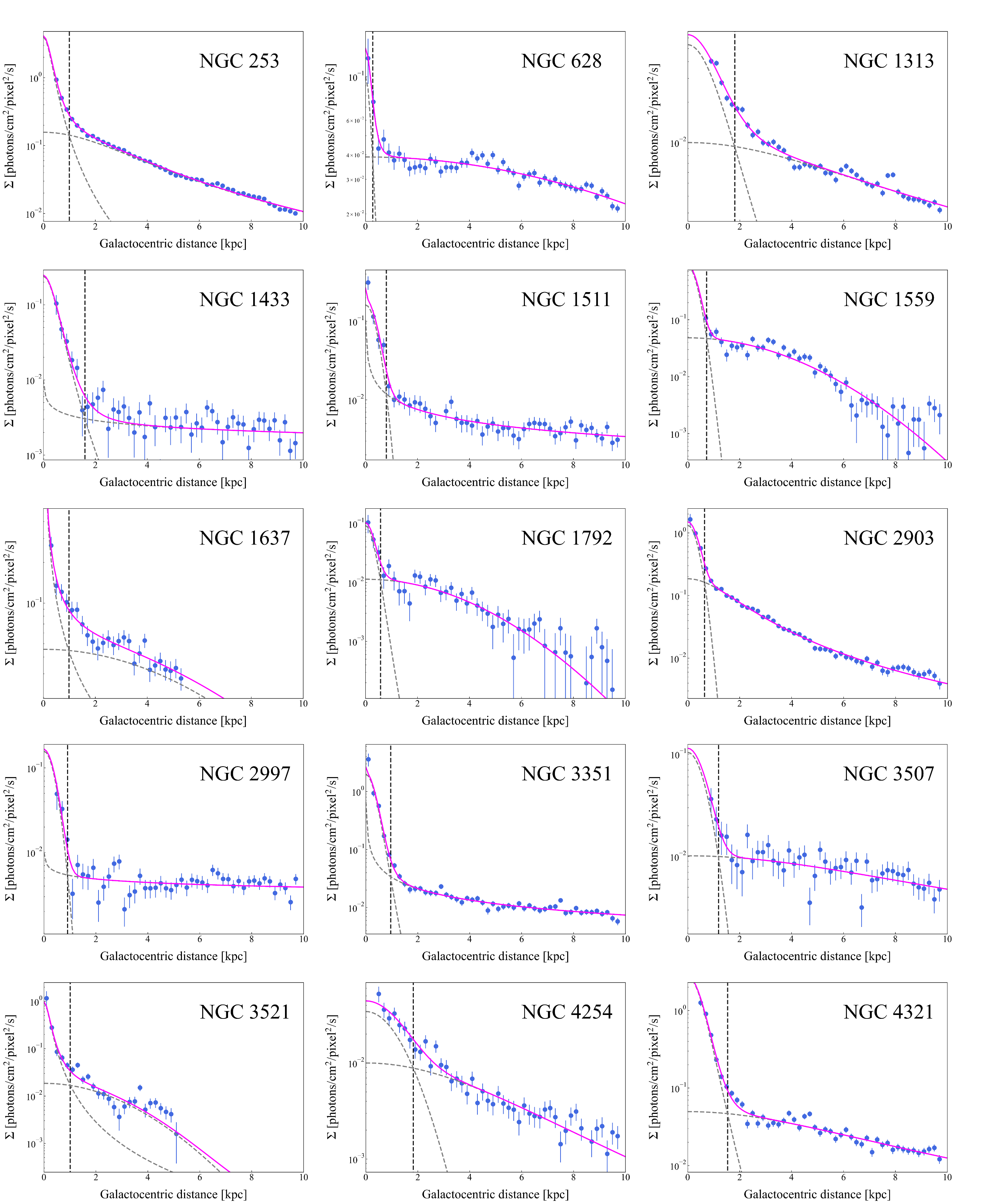}
\end{figure*}
\setcounter{figure}{0}

\begin{figure*}
\centering
\includegraphics[scale=0.5465]{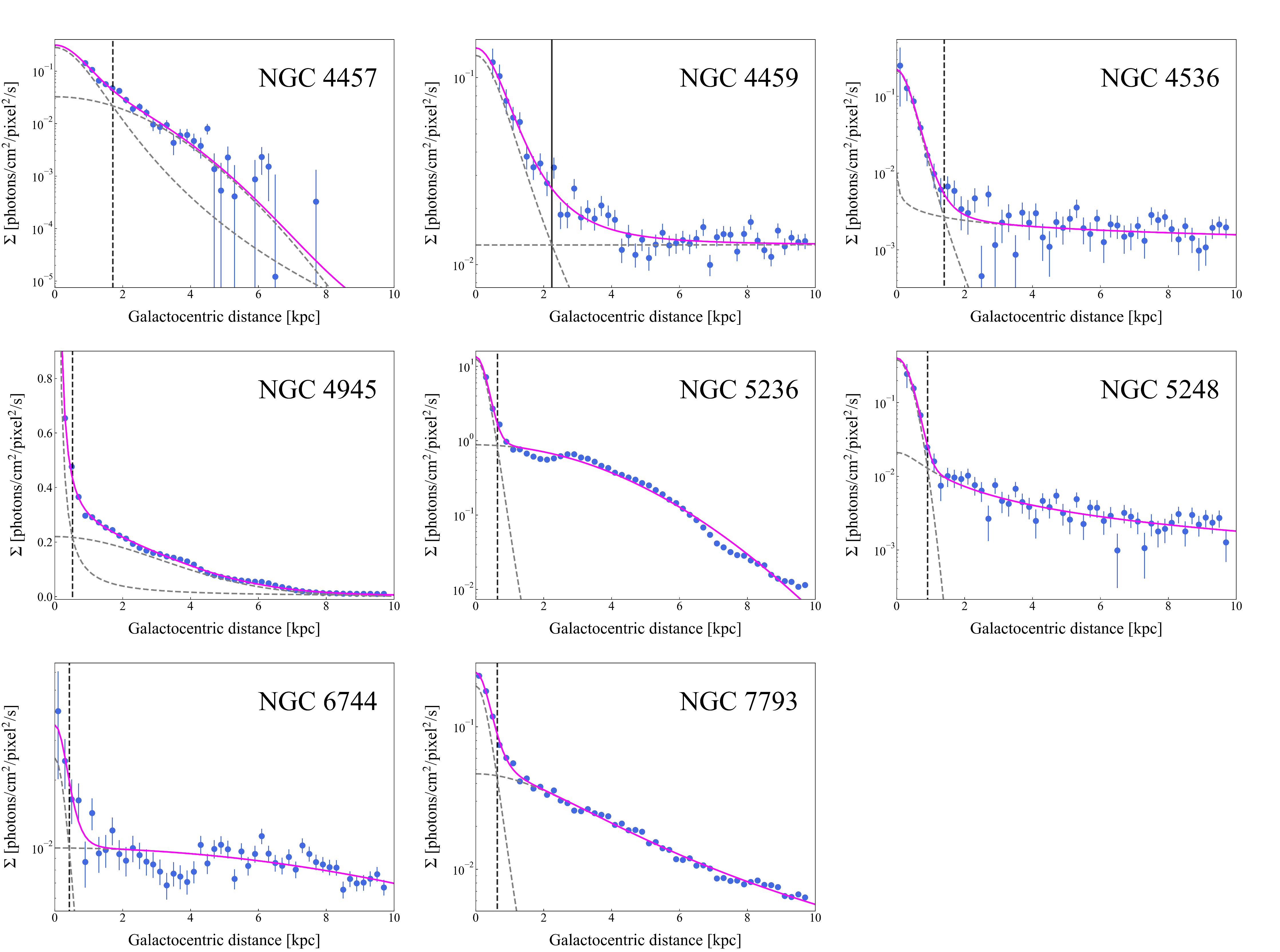}
\caption{Radial profile of the background-subtracted surface brightness in the 0.5$-$2 keV band for our sample of 23 nearby star-forming galaxies. The profile is well described ($\chi^2_{\nu}$ $\thicksim$1.4 on average) by a double $\beta$-model (magenta solid curve). The grey dashed curve represents the $\beta$-model, and the vertical black dashed line in each panel indicates the break radius of the surface brightness profile, located at the intersection of the two $\beta$-models. }
\label{fig:radial}
\end{figure*}

\emph{Spectral Fitting}. The X-ray spectra of the point source-free emission were extracted using the CIAO routine \emph{specextract}. Since all of our galaxies are small enough on the sky, the background regions are defined on the same CCD chip, covering most of the CCD area while excluding point sources and the target region. We grouped the extracted spectra in order to have minimum 10 counts per bin to apply the ${\chi}^2$ statistics. To model the spectrum of the hot gas, we used an absorbed Astrophysical Plasma Emission Code \citep[APEC;][]{2001_Smith_ApJ}, which is based on the thermal plasma emission model introduced by \citet{1977_Raymond_ApJS} but incorporates modern computing techniques and more accurate atomic data. 

The spectral fitting was performed using the XSPEC v.12.11.0 \citep{1996_Arnaud_ASPC} software. In addition to the absorbed APEC model, we also considered a two-component spectral model with a power-law plus a thermal plasma (e.g., \emph{TBABS $\times$ (APEC + POWER-LAW)}). For most galaxies with a weak X-ray emission, we fixed the photon index of the power-law component to $\Gamma$ = 1.8, which is a typically expected value for the X-ray binaries and AGNs in nearby galaxies \citep{1994_Nandra_MNRAS_405N,2002_Kong_ApJ_738K,2004_Swartz_ApJS_519S}. However, we found that the two-component spectral model does not significantly improve the fitting results or better constrain the parameters. For this reason, we selected the absorbed APEC model to calculate the hot gas luminosity in the 0.5$-$2 keV band in this work. Using the best-fit absorbing column densities, we have corrected all X-ray luminosities for both Galactic and intrinsic absorption.

\subsection{ALMA }\label{sec:alma}

The CO(2-1) data for our sample galaxies were obtained from the PHANGS-ALMA Large Program \citep[PI: E. Schinnerer;][]{2021_Leroy_ApJS,2021_Leroy_ApJS_b} at the PHANGS website\footnote{\url{https://sites.google.com/view/phangs/home/data}}. These data include both interferometric and single-dish observations, and thus provide excellent spatial resolution and sensitivity to allow us to measure various relations across a wide range of star-forming environments within the local Universe. Furthermore, the CO(2-1) data have good coverage of the areas of active star formation in each galaxy ($\thicksim$100 kpc$^2$ on average). The strategy for sample selection, observational setup, and data reduction are described in \citep{2021_Leroy_ApJS, 2021_Leroy_ApJS_b}.

In this study, we use the integrated intensity maps from the ``broad'' mask scheme, which is optimized for completeness through highly inclusive signal masking and contains the entirety of the galaxy emission. Following \citet{1997_solomon_apj}, we calculated the CO(2-1) line luminosity $L^\prime_{\rm CO}$ via 
\begin{equation}\label{eq1}
\begin{split}
L^\prime_{\rm CO}= & 3.25\times10^7 \left(\frac{S\Delta v}{{\rm 1\ Jy\ km\ s^{-1}}}\right)\left(\frac{\nu_{\rm obs}}{{\rm 1\ GHz}}\right)^{-2}\\
& \times\left(\frac{D_{\rm L}}{{\rm 1\ Mpc}}\right)^2 \left(1+z\right)^{-3}\ {\rm K\ km\ s^{-1}\ pc^2},
\end{split}
\end{equation}
where $S\Delta v$ is the velocity-integrated flux density, $\nu_{\rm obs}$ is the observed line frequency, $D_{\rm L}$ is the luminosity distance, and $z$ is the redshift. We then derived the molecular gas mass from the line luminosity $L^\prime_{\rm CO}$ via 
\begin{equation}\label{eq2}
    M_{\rm mol} = \alpha_{\rm CO(1-0)}R^{\,-1}_{21}L^\prime_{\rm CO}.
\end{equation}
Here, $\alpha_{\rm CO(1-0)}$ is the CO-to-H$_2$ conversion factor for the CO(1-0) emission line. We assume the factor to be the standard Milky Way value $\alpha_{\rm CO(1-0)}$ = 4.35 $M_\odot$ pc$^{-2}$ (K km s$^{-1}$)$^{-1}$ throughout our sample, which is commonly adopted for massive, solar-metallicity galaxies \citep{2013_Bolatto_ARA&A}. $R_{21}$ = 0.65 is the widely used CO(2-1)-to-CO(1-0) line ratio \citep[see][]{2013_Leroy_AJ_19L, 2021_den_Brok_MNRAS_3221D}.

\begin{figure*}
\centering 
\includegraphics[scale=1.01]{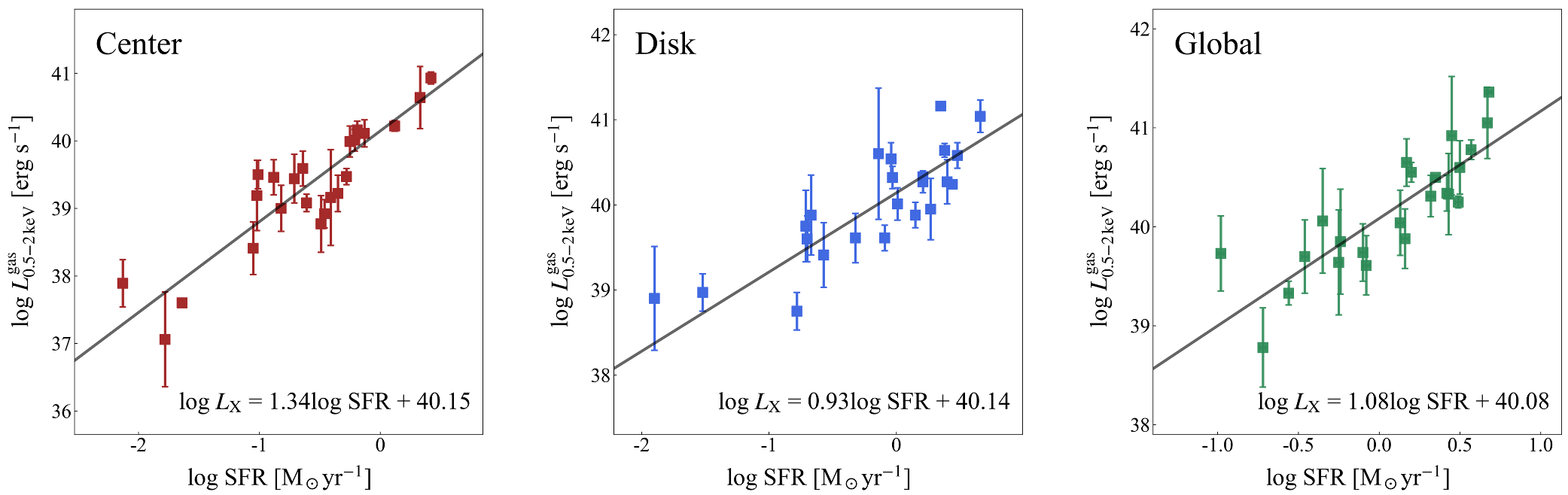}
\vspace{0.07cm}
\caption{Correlations between the diffuse X-ray emission of hot gas and the SFR in the center, disk, and on the global scale of the galaxies. The solid line in each panel shows the best-fit relation derived from the Bayesian method for linear regression. The X-ray luminosity is corrected for both Galactic and intrinsic absorption, with the error given at the 1$\sigma$ confidence level. }
\vspace{0.6cm}
\label{fig:Lx_SFR_three}
\end{figure*}

\subsection{WISE and GALEX}\label{sec:sfr}

We use near-IR imaging data from the Wide-field Infrared Survey Explorer (WISE) to trace the stellar mass. The WISE W1 band (\mbox{3.4 $\mu$m}) images are compiled by the z0MGS project \citep{2019_Leroy_ApJS_24L}. All these data have been background-subtracted, foreground star-masked, and convolved from a non-Gaussian PSF to a 7$\arcsec$.5 Gaussian PSF using the convolution kernels provided by \citet{2011_Aniano_Kernels}. We convert the WISE \mbox{3.4 $\mu$m} intensity to stellar mass surface density $\Sigma_{\rm \star}$ via
\begin{equation}\label{eq3}
    \frac{\Sigma_{\rm \star}}{M_\odot\,{\rm pc}^{-2}} = 330 \left(\frac{\Upsilon_{3.4\,\mu m}}{0.5}\right)\left(\frac{I_{3.4\,\mu m}}{\rm MJy~sr^{-1}}\right) {\rm cos}\,i.
\end{equation}
The cos\,$i$ factor accounts for the inclination of the sample galaxies and $\Upsilon_{3.4\,\mu m}$ is the stellar mass-to-light ratio at \mbox{3.4 $\mu$m}. We adopt a fixed mass-to-light ratio of $\Upsilon_{3.4\,\mu m}$ = 0.35 $M_\odot$ $L_\odot^{-1}$, as our sample consists entirely of star-forming galaxies and only WISE data are used to estimate the stellar mass \citep{2019_Leroy_ApJS_24L,2021_Leroy_ApJS}.

\begin{deluxetable}{lcccc}
\tablecaption{Best-fit Parameters for the \lx$-$SFR relations \label{tab:table2}}
\addtolength{\tabcolsep}{3.0pt}
\tablewidth{0.5pt}
\tablehead{
\colhead{Region} & \colhead{$\alpha $} & \colhead{$\beta $} & \colhead{$\sigma $} & \colhead{$r_{\rm s}$} \\ 
\colhead{(1)} & \colhead{(2)} & \colhead{(3)} & \colhead{(4)} & \colhead{(5)}
}
\startdata
Center & 1.34 $\pm$ 0.16 & 40.15 $\pm$ 0.13 & 0.41 & 0.83 \\
Disk & 0.93 $\pm$ 0.16 & 40.14 $\pm$ 0.09 & 0.36 & 0.79 \\
Global & 1.08 $\pm$ 0.17 & 40.08 $\pm$ 0.08 & 0.31 & 0.88 \\
\cline{1-5}
0.5 kpc & 1.38 $\pm$ 0.21 & 40.16 $\pm$ 0.22 & 0.38 & 0.84 \\
1 kpc & 1.33 $\pm$ 0.17 & 40.14 $\pm$ 0.13 & 0.37 & 0.89 \\
1.5 kpc & 1.25 $\pm$ 0.17 & 40.11 $\pm$ 0.10 & 0.37 & 0.87 \\
2 kpc & 1.16 $\pm$ 0.18 & 40.15 $\pm$ 0.10 & 0.35 & 0.89 \\
2.5 kpc & 1.10 $\pm$ 0.18 & 40.17 $\pm$ 0.09 & 0.30 & 0.87 \\
3 kpc & 1.11 $\pm$ 0.19 & 40.18 $\pm$ 0.08 & 0.28 & 0.88 \\
\enddata
\tablecomments{ $\alpha$ and $\beta$ are the slope and intercept of the scaling relation between {\lx} and SFR. $\sigma$ is the $rms$ scatter and $r_{\rm s}$ is the Spearman rank correlation coefficient.  }
\vspace{-0.7cm}
\end{deluxetable}

We combine mid-IR images from WISE and far-/near-UV images from the Galaxy Evolution Explorer (GALEX) to estimate the star formation rate. These data also come from the z0MGS project \citep{2019_Leroy_ApJS_24L} and have been convolved with the convolution kernels provided by \citet{2011_Aniano_Kernels} to a 15$\arcsec$ Gaussian PSF. By default, we adopt the prescription described by \citet{2019_Leroy_ApJS_24L}, using the combination of GALEX FUV (154 nm) and WISE \mbox{22 $\mu$m} data to calculate the SFR. Since NGC 1559 does not have FUV data, we instead use a combination of GALEX NUV (231 nm) and WISE \mbox{22 $\mu$m} data 
\begin{equation}\label{eq4}
    \frac{\rm SFR}{M_\odot\,{\rm yr}^{-1}} = 10^{-43.42} \times (\nu L_{\nu})_{\rm FUV} + 10^{-42.73} \times (\nu L_{\nu})_{22 \mu m},
\end{equation}
\begin{equation}\label{eq5}
    \frac{\rm SFR}{M_\odot\,{\rm yr}^{-1}} = 10^{-43.24} \times (\nu L_{\nu})_{\rm NUV} + 10^{-42.79} \times (\nu L_{\nu})_{22 \mu m}.
\end{equation}
where $\nu L_{\nu}$ is the luminosity in a given band (UV or \mbox{22 $\mu$m}) in units of erg s$^{-1}$ and measured as 4$\pi$D$^2_{\rm L}$$\nu f_{\nu}$. For NGC 1637 and NGC 4945, where UV data are not available, we use WISE data alone
\begin{equation}\label{eq6}
    \frac{\rm SFR}{M_\odot\,{\rm yr}^{-1}} = 10^{-42.63} \times (\nu L_{\nu})_{22 \mu m}.
\end{equation}

Since the galaxies in our sample are all nearby ($\textless$ 24 Mpc), the 15$\arcsec$ PSF of the WISE and GALEX data is typically smaller than the regions studied in this work, even for the smallest scales of $\thicksim$0.5 kpc, and thus has little effect on the estimation of the SFR.

\section{Results}\label{sec:results}

\subsection{Surface Brightness Profile of Hot Gas}\label{Radial_profile}

We used the point source-excluded maps to extract the surface brightness profiles of our 23 sample galaxies in the 0.5$-$2 keV band. The profiles span a spatial scale of 20 kpc for each galaxy, and the interval of the radial bins is 0.2 kpc. Figure~\ref{fig:radial} shows the background-subtracted surface brightness profiles fitted with a two-component $\beta$-model. The $\beta$-model is a convenient analytical expression widely used in studies of galaxies and clusters, described as 
\begin{equation}\label{eq7}
    \Sigma(r)=\Sigma_{0}\left[1+\left(\frac{r}{r_{\mathrm{0}}}\right)^{2}\right]^{-3 \beta+0.5}.
\end{equation}
where $\Sigma_{0}$ is the central surface brightness, $r_{\mathrm{0}}$ is the scale radius, and $\beta$ is the power-law index. The profiles of our sample galaxies are well fitted by the double $\beta$-model with a reduced $\chi^2$ $\thicksim$ 1.4 on average.

From Figure~\ref{fig:radial}, we can determine the break points in the surface brightness profiles, which are defined as the intersection points of the two $\beta$-models. The break radius, $r_{\rm c}$, in the profile is marked by a vertical dashed line in Figure~\ref{fig:radial}, and its value is presented in Table~\ref{tab:table1}. As seen, the break points determined from the fits are broadly consistent with estimates from visual inspection of the profile morphology, where a sharp radial decrease in surface brightness occurs within the inner region and a slow, extended variation is seen outside it. We infer that the break indicates different origins or heating mechanisms for the hot gas inside and outside the break radius. Therefore, we refer to the region within the break radius as the ``center'' of our sample galaxies and the region outside this radius as the ``disk'' hereafter. As shown in Table~\ref{tab:table1}, we find that the break radii of these galaxies are significantly different, ranging from roughly 0.3 to 2 kpc, with an average of 1.1 kpc.

\begin{figure}
\centering 
\includegraphics[scale=0.515]{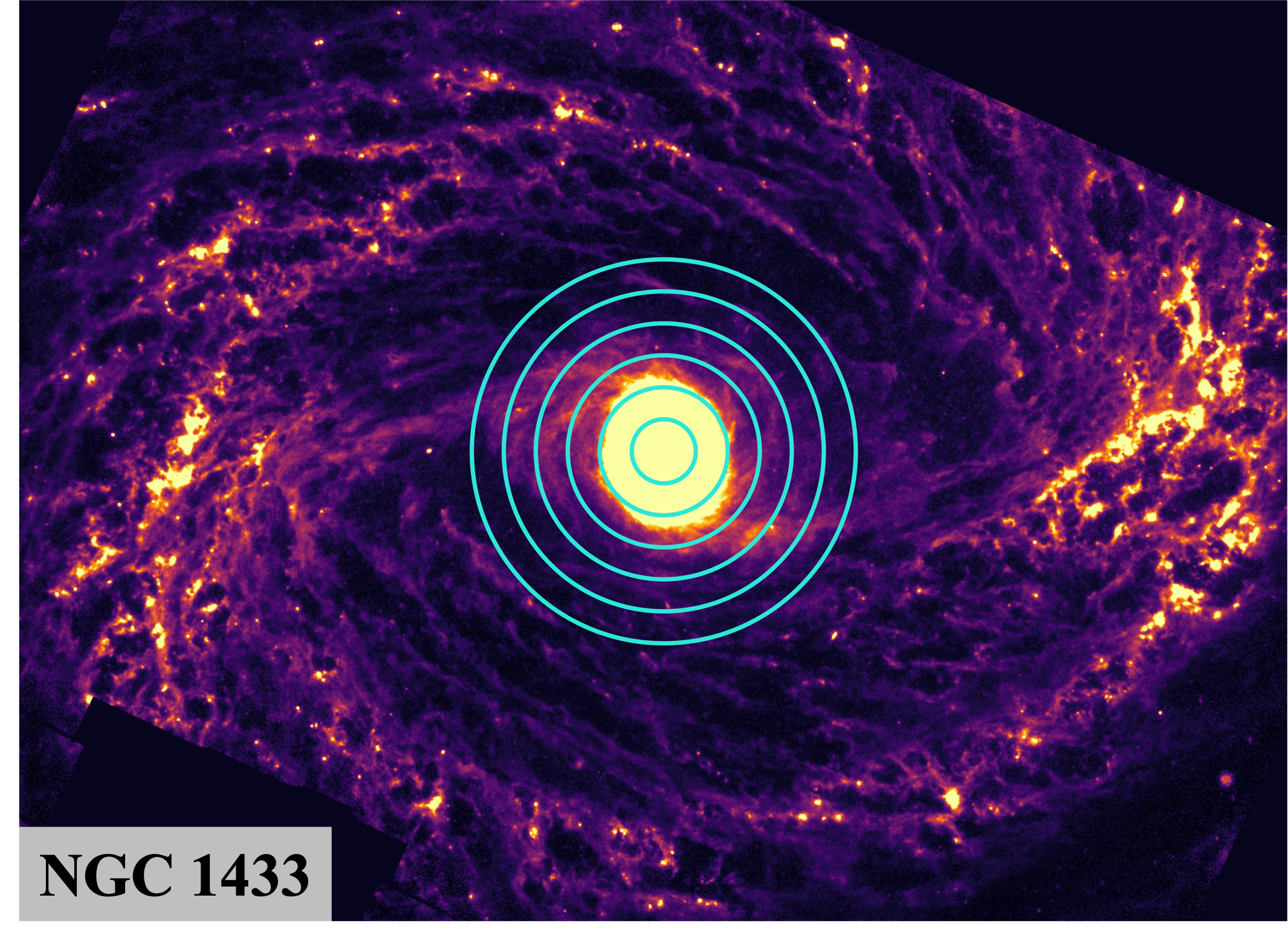}
\caption{Example of JWST MIRI F1130W map for one of the galaxies in our sample (NGC 1433). The cyan concentric circles correspond to physical areas with radii of 0.5, 1, 1.5, 2, 2.5, and 3 kpc.  }
\vspace{-0.0cm}
\label{fig:n1433}
\end{figure}

\subsection{Correlation between SFR and Hot Gas Luminosity}\label{sec:Lx-LIR}

Figure~\ref{fig:Lx_SFR_three} presents the correlations between the SFR and the diffuse hot gas luminosity in the 0.5$-$2 keV band for different regions (e.g., the centers, disks, and entire galaxies; see Section~\ref{Radial_profile} for the definitions of the center and disk) of our 23 sample galaxies. The thermal X-ray emission is strongly correlated with the SFR, spanning three orders of magnitude in SFR and five orders of magnitude in X-ray luminosity, with Spearman rank correlation coefficients of 0.83, 0.79, and 0.88, respectively (see Table~\ref{tab:table2}). We fitted the {\lx}$-$SFR relations with a linear model, ${\rm log}$$L_{\rm 0.5-2\,keV}^{\rm gas}$ = ${\rm \alpha}$\,${\rm log}$${\rm SFR}$ $+$ ${\rm \beta}$, using the Bayesian linear regression method implemented in the IDL routine LINMIX\_ERR \citep{2007_Kelly_Apj}. The following are the best-fit relations with uncertainties
\begin{equation}\label{eq8}
\begin{split}
Center:&\, {\rm log}\,(\frac{L_{\rm 0.5-2\,keV}^{\rm gas}} {\rm erg\ s^{-1}}) = \\ & 1.34(\pm0.16) {\rm log}\,\frac{\rm SFR} {M_\odot\ {\rm yr}^{-1}} + 40.15(\pm0.13),  
\end{split}
\end{equation}
\begin{equation}\label{eq9}
\begin{split}
Disk:&\, {\rm log}\,(\frac{L_{\rm 0.5-2\,keV}^{\rm gas}} {\rm erg\ s^{-1}}) = \\ & 0.93(\pm0.16) {\rm log}\,\frac{\rm SFR} {M_\odot\ {\rm yr}^{-1}} + 40.14(\pm0.09),  
\end{split}
\end{equation}
\begin{equation}\label{eq10}
\begin{split}
Global:&\, {\rm log}\,(\frac{L_{\rm 0.5-2\,keV}^{\rm gas}} {\rm erg\ s^{-1}}) = \\ & 1.08(\pm0.17) {\rm log}\,\frac{\rm SFR} {M_\odot\ {\rm yr}^{-1}} + 40.08(\pm0.08).  
\end{split}
\end{equation}

\begin{figure*}
\centering 
\includegraphics[scale=1.02]{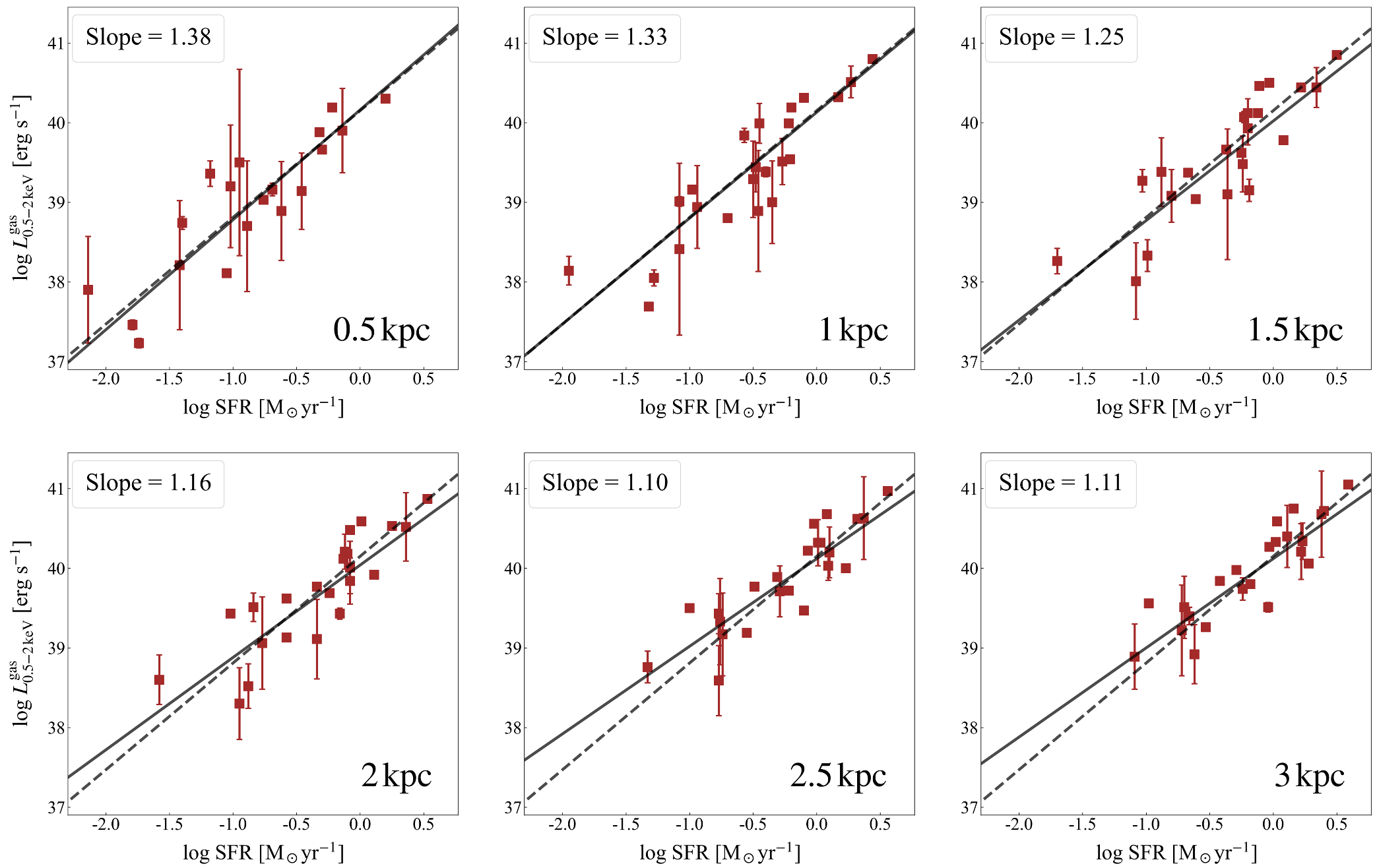}
\vspace{-0.5cm}
\caption{Correlations between diffuse X-ray luminosity and SFR in the galactic central regions with radii of 0.5, 1, 1.5, 2, 2.5, and 3 kpc. For the region with a radius of 0.5 kpc (upper left panel), four galaxies are excluded due to their very low number of counts in the X-ray maps after subtracting the bright point sources (see table~\ref{tab:table3}). The solid lines indicate the best-fit relations by the Bayesian method and the dashed lines represent the relation of the center (Equation~\ref{eq8}). The best-fit slope is shown in the top left corner of each panel. } 
\vspace{0.5cm}
\label{fig:multi_Lx_SFR}
\end{figure*}

These relations indicate that the bulk of the diffuse X-ray emission is indeed the product of star formation but the thermal luminosity per unit of star formation rate varies in different regions of these galaxies. For the entire galaxy, we calculated the luminosity of the region with a radius of 1.5$R_{\rm e}$ as the total X-ray emission of the galaxy. The slope of the global {\lx}$-$SFR relation is slightly higher than unity, which is consistent with previous studies \citep[e.g.,][]{2003_Ranalli_A&A,2005_Grimes_ApJ_187G,2009_Owen_MNRAS_1741O,2012_Mineo_hotgas,2013_LiJiangTao_MNRAS_2}. For the disk, the slope is shallower than unity, and it is similar to the result of \citet{2025_Zhangcy_ApJ_15Z}, who studied the spiral galaxy M51 at a 1.3 kpc scale and found a slope of 0.88 for the diffuse X-ray luminosity binned by total-IR (an indicator for SFR) after excluding the central 2 kpc region of M51. \citet{2020_Kouroumpatzakis_mnras} divided 13 resolved star-forming galaxies using a set of subgalactic-scale grids (e.g., 1$\times$1, 2$\times$2, 3$\times$3, and 4$\times$4 kpc$^2$) and also found the sub-linear relations between multi-band X-ray luminosity and SFR, with the slope increasing with spatial scale until approaching one. From this perspective, we can understand why the slope of the disk in our work is slightly less than unity. For the center, a super-linear relation is between the diffuse X-ray luminosity and the SFR. This is due to the fact that core-collapse supernovae are the primary source of energy for heating gas in the galactic center, and their X-ray radiation efficiency (e.g., {\lx}/$\dot{E}_\mathrm{SN}$, where $\dot{E}_\mathrm{SN}$ is the supernova mechanical energy input rate) increases with the SFR \citep{2025_Zhangcy_ApJ_15Z}. More details on X-ray radiation efficiency and stellar feedback are discussed in Section~\ref{sec:Efficiency}.

\begin{figure}
\centering 
\includegraphics[scale=0.51]{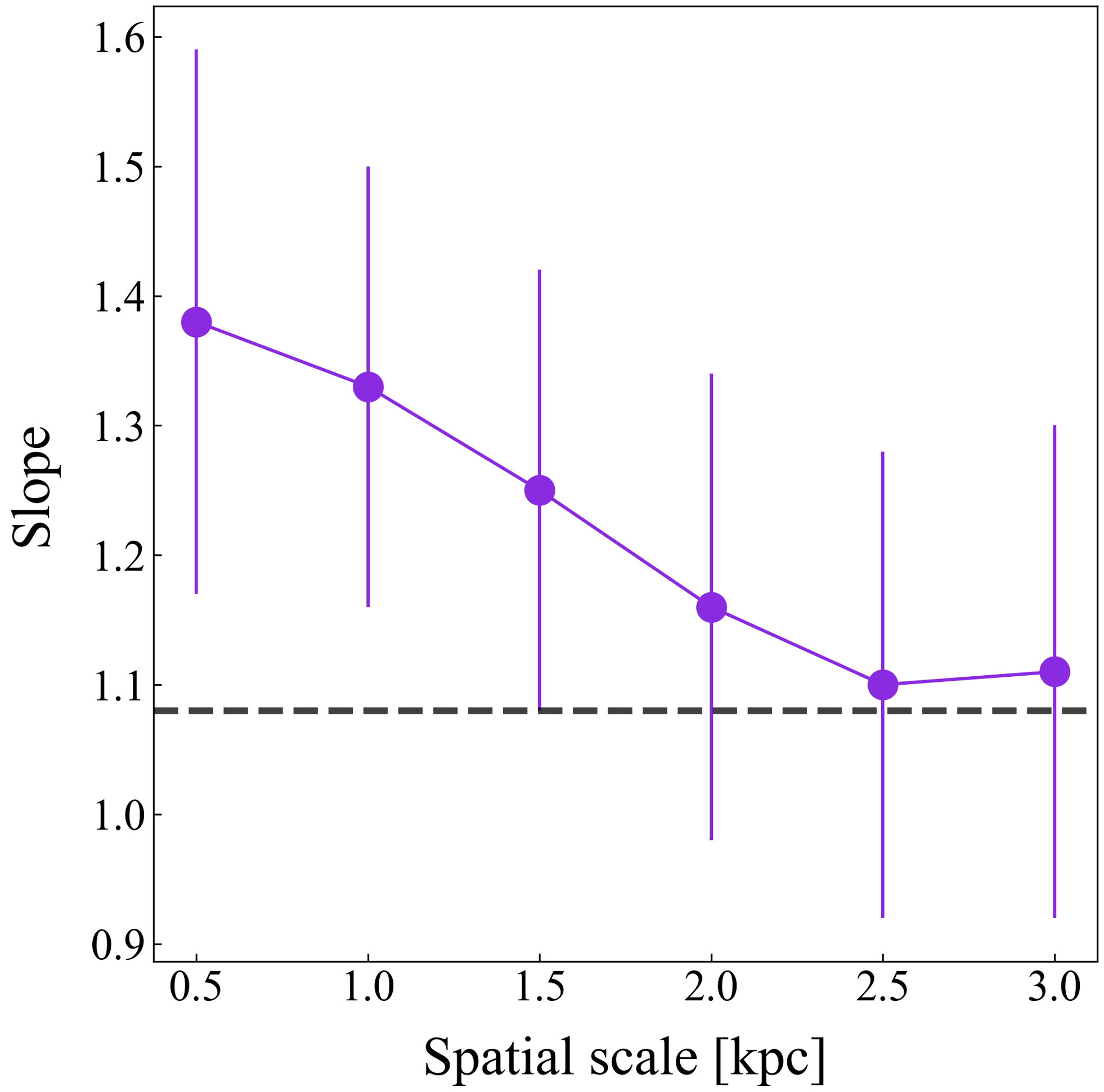}
\vspace{-0.5cm}
\caption{The slope of the relation between central diffuse X-ray luminosity and SFR as a function of spatial scale. The $x$-axis denotes the radius of the central region of our sample galaxies. The black dashed line is the slope of the global {\lx}$-$SFR relation (e.g., Equation~\ref{eq10}). } 
\vspace{0.1cm}
\label{fig:slope_scale}
\end{figure}

\subsection{Variation of Slope at Multiple Spatial Scales}\label{sec:spatial_scales}

From Section~\ref{sec:Lx-LIR}, we have known that the slope of the $L_{\rm X}$$-$SFR relation increases with the spatial scale in the galactic disk \citep{2020_Kouroumpatzakis_mnras}. In this section, we explore how varying the spatial scale of the galactic central region affects the {\lx}$-$SFR relation. We defined concentric circles with radii of 0.5, 1, 1.5, 2, 2.5, and 3 kpc centered on the positions of our sample galaxies. The minimum radius of 0.5 kpc was limited by the counts of the $Chandra$ data, as below this physical scale most galaxies in the sample are difficult to detect with sufficient number of photons to estimate the diffuse X-ray luminosity. Figure~\ref{fig:n1433} shows a schematic representation of the concentric circles for an example galaxy NGC 1433. The SFR and the hot gas luminosity in the 0.5$-$2 keV band for each circular region are listed in Table~\ref{tab:table3}.  

Figure~\ref{fig:multi_Lx_SFR} shows the {\lx}$-$SFR relation on the multiple spatial scales in the galactic central region. The measured slopes and intercepts are listed in Table~\ref{tab:table2}. At all spatial scales, the {\lx} and SFR show a strong correlation, and the Spearman rank correlation coefficients are all higher than 0.8 (see Table~\ref{tab:table2}). We plotted Equation~\ref{eq8} in Figure~\ref{fig:multi_Lx_SFR} as a dashed line to present the variation of the scaling relation with the spatial scale and found that the relation at the 1 kpc scale is the closest to Equation~\ref{eq8}. 

Figure~\ref{fig:slope_scale} presents the slope of the {\lx}$-$SFR relations as a function of the spatial scale. In contrast to the galactic disk, the slope shows a negative correlation with spatial scale in the central regions of the sample galaxies. Until it declines at 2.5 kpc, the slope levels off and becomes consistent with the global result. This suggests that the characteristics of hot gas in the central regions of galaxies are almost completely erased on a scale of $\thicksim$5 kpc.

The $rms$ scatter, $\sigma$, of the {\lx}$-$SFR relation is also systematically lower for larger spatial scales, consistent with the previous findings \citep[e.g.,][]{2008_Bigiel_AJ_2846B,2013_Leroy_AJ_19L,2020_Kouroumpatzakis_mnras,2021_Pessa_A&A_134P}. We argue that the lower scatter at larger spatial scales is due to the averaging of small-scale variations of regions in different physical phases of their evolutionary cycles.

\begin{deluxetable*}{l|cc|cc|cc|cc|cc|cc}
\centering
\tablecaption{Central Hot Gas Luminosity and SFR at Different Spatial Scales \label{tab:table3}}
\addtolength{\tabcolsep}{0pt}
\tablewidth{0pt}
\tablehead{
\colhead{} & \multicolumn{2}{c}{0.5 kpc} & \multicolumn{2}{c}{1 kpc} & \multicolumn{2}{c}{1.5 kpc} & \multicolumn{2}{c}{2 kpc} & \multicolumn{2}{c}{2.5 kpc} & \multicolumn{2}{c}{3 kpc} \\
\colhead{Galaxy} & \colhead{\lx} & \colhead{SFR} & \colhead{\lx} & \colhead{SFR} & \colhead{\lx} & \colhead{SFR} & \colhead{\lx} & \colhead{SFR} & \colhead{\lx} & \colhead{SFR} & \colhead{\lx} & \colhead{SFR} 
}
\startdata
NGC$\,$253 & 2.0$\,$$\times$$\,$10$^{40}$ & 2.40 & 6.3$\,$$\times$$\,$10$^{40}$ & 2.74 & 7.1$\,$$\times$$\,$10$^{40}$ & 3.17 & 7.4$\,$$\times$$\,$10$^{40}$ & 3.39 & 9.3$\,$$\times$$\,$10$^{40}$ & 3.66 & 1.1$\,$$\times$$\,$10$^{41}$ & 3.93 \\
NGC$\,$628 & 1.7$\,$$\times$$\,$10$^{37}$ & 0.02 & 4.9$\,$$\times$$\,$10$^{37}$ & 0.05 & 1.0$\,$$\times$$\,$10$^{38}$ & 0.08 & 2.0$\,$$\times$$\,$10$^{38}$ & 0.11 & 3.9$\,$$\times$$\,$10$^{38}$ & 0.17 & 8.3$\,$$\times$$\,$10$^{38}$ & 0.24 \\
NGC$\,$1313 & 5.5$\,$$\times$$\,$10$^{38}$ & 0.04 & 1.4$\,$$\times$$\,$10$^{39}$ & 0.10 & 2.4$\,$$\times$$\,$10$^{39}$ & 0.21 & 4.2$\,$$\times$$\,$10$^{39}$ & 0.27 & 5.9$\,$$\times$$\,$10$^{39}$ & 0.33 & 6.9$\,$$\times$$\,$10$^{39}$ & 0.38 \\
NGC$\,$1433 & ... & ... & 8.8$\,$$\times$$\,$10$^{38}$ & 0.12 & 1.2$\,$$\times$$\,$10$^{39}$ & 0.16 & 1.2$\,$$\times$$\,$10$^{39}$ & 0.17 & 1.5$\,$$\times$$\,$10$^{39}$ & 0.18 & 1.7$\,$$\times$$\,$10$^{39}$ & 0.19 \\
NGC$\,$1511 & 1.5$\,$$\times$$\,$10$^{39}$ & 0.20 & 3.4$\,$$\times$$\,$10$^{39}$ & 0.62 & 6.1$\,$$\times$$\,$10$^{39}$ & 1.21 & 8.3$\,$$\times$$\,$10$^{39}$ & 1.30 & 1.0$\,$$\times$$\,$10$^{40}$ & 1.69 & 1.2$\,$$\times$$\,$10$^{40}$ & 1.93 \\
NGC$\,$1559 & ... & ... & 2.0$\,$$\times$$\,$10$^{39}$ & 0.31 & 8.5$\,$$\times$$\,$10$^{39}$ & 0.64 & 1.0$\,$$\times$$\,$10$^{40}$ & 0.83 & 1.6$\,$$\times$$\,$10$^{40}$ & 1.26 & 2.2$\,$$\times$$\,$10$^{40}$ & 1.68 \\
NGC$\,$1637 & 7.8$\,$$\times$$\,$10$^{38}$ & 0.24 & 2.4$\,$$\times$$\,$10$^{39}$ & 0.39 & 4.6$\,$$\times$$\,$10$^{39}$ & 0.42 & 5.9$\,$$\times$$\,$10$^{39}$ & 0.46 & 7.8$\,$$\times$$\,$10$^{39}$ & 0.49 & 9.6$\,$$\times$$\,$10$^{39}$ & 0.52 \\
NGC$\,$1792 & 1.6$\,$$\times$$\,$10$^{39}$ & 0.10 & 2.8$\,$$\times$$\,$10$^{39}$ & 0.33 & 3.1$\,$$\times$$\,$10$^{39}$ & 0.57 & 6.9$\,$$\times$$\,$10$^{39}$ & 0.83 & 1.1$\,$$\times$$\,$10$^{40}$ & 1.22 & 1.6$\,$$\times$$\,$10$^{40}$ & 1.68 \\
NGC$\,$2903 & 7.6$\,$$\times$$\,$10$^{39}$ & 0.47 & 2.0$\,$$\times$$\,$10$^{40}$ & 0.79 & 3.2$\,$$\times$$\,$10$^{40}$ & 0.94 & 3.9$\,$$\times$$\,$10$^{40}$ & 1.02 & 4.8$\,$$\times$$\,$10$^{40}$ & 1.21 & 5.6$\,$$\times$$\,$10$^{40}$ & 1.44 \\
NGC$\,$2997 & ... & ... & 7.7$\,$$\times$$\,$10$^{38}$ & 0.35 & 1.3$\,$$\times$$\,$10$^{39}$ & 0.44 & 1.3$\,$$\times$$\,$10$^{39}$ & 0.46 & 5.1$\,$$\times$$\,$10$^{39}$ & 0.51 & 5.5$\,$$\times$$\,$10$^{39}$ & 0.58 \\
NGC$\,$3351 & 1.4$\,$$\times$$\,$10$^{39}$ & 0.34 & 3.2$\,$$\times$$\,$10$^{39}$ & 0.53 & 4.1$\,$$\times$$\,$10$^{39}$ & 0.56 & 4.9$\,$$\times$$\,$10$^{39}$ & 0.57 & 5.3$\,$$\times$$\,$10$^{39}$ & 0.60 & 6.3$\,$$\times$$\,$10$^{39}$ & 0.65 \\
NGC$\,$3507 & ... & ... & 2.6$\,$$\times$$\,$10$^{38}$ & 0.08 & 2.4$\,$$\times$$\,$10$^{39}$ & 0.13 & 3.2$\,$$\times$$\,$10$^{39}$ & 0.14 & 2.1$\,$$\times$$\,$10$^{39}$ & 0.17 & 2.5$\,$$\times$$\,$10$^{39}$ & 0.22 \\
NGC$\,$3521 & 2.3$\,$$\times$$\,$10$^{39}$ & 0.07 & 6.9$\,$$\times$$\,$10$^{39}$ & 0.27 & 1.2$\,$$\times$$\,$10$^{40}$ & 0.59 & 1.3$\,$$\times$$\,$10$^{40}$ & 0.74 & 2.1$\,$$\times$$\,$10$^{40}$ & 1.08 & 2.1$\,$$\times$$\,$10$^{40}$ & 1.04 \\
NGC$\,$4254 & 3.1$\,$$\times$$\,$10$^{39}$ & 0.11 & 9.7$\,$$\times$$\,$10$^{39}$ & 0.36 & 1.3$\,$$\times$$\,$10$^{40}$ & 0.64 & 1.6$\,$$\times$$\,$10$^{40}$ & 0.76 & 2.1$\,$$\times$$\,$10$^{40}$ & 1.03 & 2.5$\,$$\times$$\,$10$^{40}$ & 1.30 \\
NGC$\,$4321 & 1.1$\,$$\times$$\,$10$^{39}$ & 0.17 & 9.9$\,$$\times$$\,$10$^{39}$ & 0.60 & 1.3$\,$$\times$$\,$10$^{40}$ & 0.75 & 1.5$\,$$\times$$\,$10$^{40}$ & 0.80 & 1.7$\,$$\times$$\,$10$^{40}$ & 0.86 & 1.9$\,$$\times$$\,$10$^{40}$ & 0.93 \\
NGC$\,$4457 & 1.3$\,$$\times$$\,$10$^{38}$ & 0.09 & 6.3$\,$$\times$$\,$10$^{38}$ & 0.20 & 1.1$\,$$\times$$\,$10$^{39}$ & 0.25 & 1.4$\,$$\times$$\,$10$^{39}$ & 0.26 & 1.6$\,$$\times$$\,$10$^{39}$ & 0.28 & 1.8$\,$$\times$$\,$10$^{39}$ & 0.30 \\
NGC$\,$4459 & 1.6$\,$$\times$$\,$10$^{38}$ & 0.04 & 1.0$\,$$\times$$\,$10$^{39}$ & 0.08 & 1.9$\,$$\times$$\,$10$^{39}$ & 0.09 & 2.7$\,$$\times$$\,$10$^{39}$ & 0.10 & 3.2$\,$$\times$$\,$10$^{39}$ & 0.10 & 3.6$\,$$\times$$\,$10$^{39}$ & 0.10 \\
NGC$\,$4536 & 8.0$\,$$\times$$\,$10$^{39}$ & 0.72 & 3.2$\,$$\times$$\,$10$^{40}$ & 1.88 & 2.8$\,$$\times$$\,$10$^{40}$ & 2.18 & 3.3$\,$$\times$$\,$10$^{40}$ & 2.27 & 4.3$\,$$\times$$\,$10$^{40}$ & 2.34 & 4.8$\,$$\times$$\,$10$^{40}$ & 2.40 \\
NGC$\,$4945 & 4.6$\,$$\times$$\,$10$^{39}$ & 0.50 & 1.5$\,$$\times$$\,$10$^{40}$ & 0.63 & 2.9$\,$$\times$$\,$10$^{40}$ & 0.78 & 3.0$\,$$\times$$\,$10$^{40}$ & 0.83 & 3.6$\,$$\times$$\,$10$^{40}$ & 0.96 & 3.9$\,$$\times$$\,$10$^{40}$ & 1.07 \\
NGC$\,$5236 & 1.5$\,$$\times$$\,$10$^{40}$ & 1.29 & 2.1$\,$$\times$$\,$10$^{40}$ & 1.48 & 2.7$\,$$\times$$\,$10$^{40}$ & 1.68 & 3.4$\,$$\times$$\,$10$^{40}$ & 1.76 & 4.2$\,$$\times$$\,$10$^{40}$ & 2.08 & 5.3$\,$$\times$$\,$10$^{40}$ & 2.54 \\
NGC$\,$5248 & 5.0$\,$$\times$$\,$10$^{38}$ & 0.13 & 1.0$\,$$\times$$\,$10$^{39}$ & 0.45 & 1.4$\,$$\times$$\,$10$^{39}$ & 0.64 & 2.7$\,$$\times$$\,$10$^{39}$ & 0.69 & 3.0$\,$$\times$$\,$10$^{39}$ & 0.80 & 3.2$\,$$\times$$\,$10$^{39}$ & 0.91 \\
NGC$\,$6744 & 8.0$\,$$\times$$\,$10$^{37}$ & 0.01 & 1.4$\,$$\times$$\,$10$^{38}$ & 0.01 & 1.8$\,$$\times$$\,$10$^{38}$ & 0.02 & 4.0$\,$$\times$$\,$10$^{38}$ & 0.03 & 5.8$\,$$\times$$\,$10$^{38}$ & 0.05 & 7.8$\,$$\times$$\,$10$^{38}$ & 0.08 \\
NGC$\,$7793 & 2.9$\,$$\times$$\,$10$^{37}$ & 0.02 & 1.1$\,$$\times$$\,$10$^{38}$ & 0.05 & 2.1$\,$$\times$$\,$10$^{38}$ & 0.10 & 3.3$\,$$\times$$\,$10$^{38}$ & 0.13 & 2.7$\,$$\times$$\,$10$^{39}$ & 0.17 & 3.2$\,$$\times$$\,$10$^{39}$ & 0.20 \\
\enddata 
\tablecomments{ {\lx} in units of erg s$^{\rm -1}$; SFR in units of $M_{\odot}$ ${\rm yr}^{-1}$. }
\vspace{-0.3cm}
\end{deluxetable*}

\section{Discussion} \label{sec:discussion}

In previous sections, we presented the scaling relation between {\lx} and SFR in different regions of galaxies, and we found that the slope of this relation anticorrelates with spatial scale in the galactic central region. In the following, we focus on the effects of different physical parameters on the hot gas in the center of the galaxy. We first discuss the correlation of stellar mass with the thermal X-ray emission, and investigate whether stellar mass affects the {\lx}$-$SFR relation. Then we explore the role of molecular gas and baryonic mass. The contamination of faint compact sources in the soft X-ray band is discussed at the final section.

\begin{figure*}
\centering 
\includegraphics[scale=0.99]{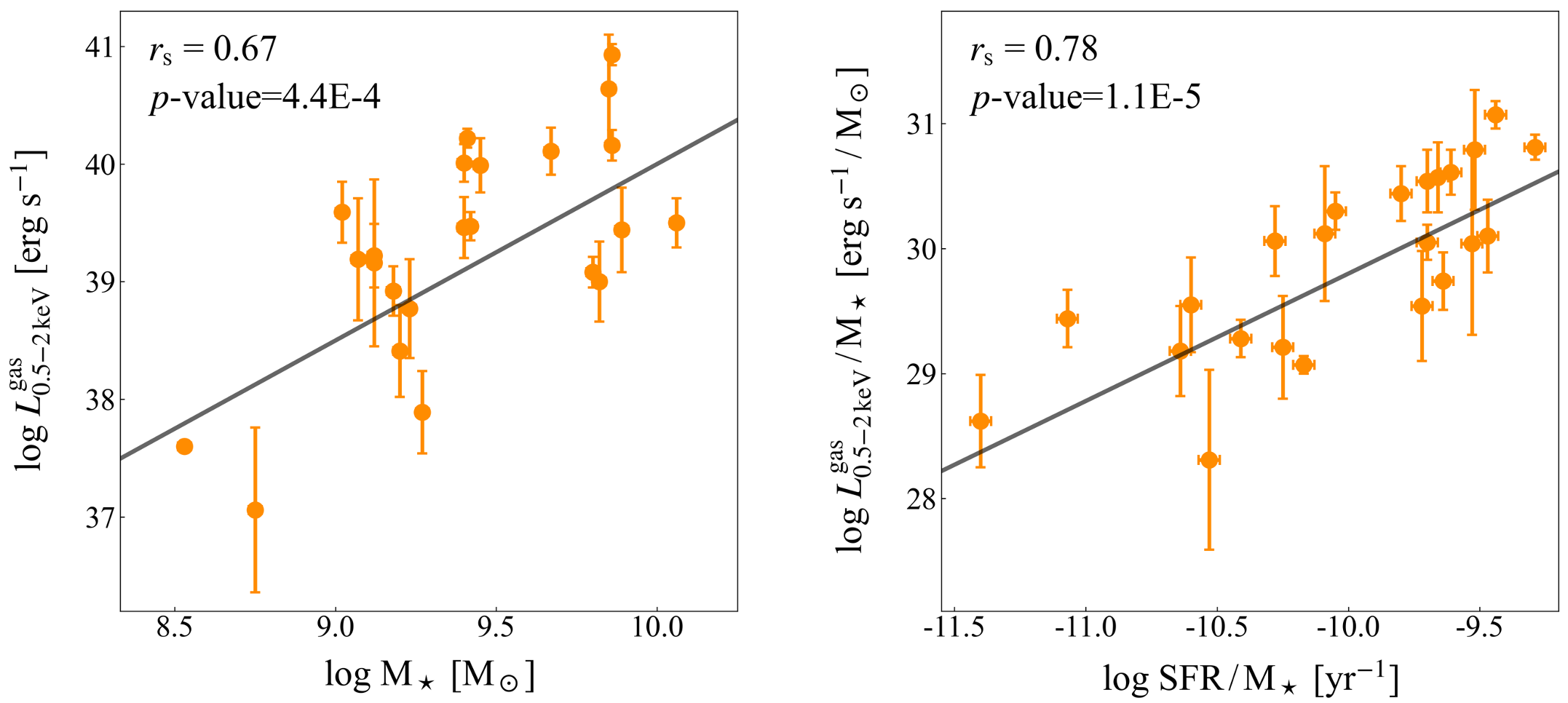}
\vspace{-0.3cm}
\caption{The diffuse X-ray luminosity as a function of stellar mass ($left$) and {\lx}/$M_{\rm \star}$ as a function of sSFR ($right$) for the centers of the galaxies. The X-ray luminosity and SFR are normalized by $M_{\rm \star}$ in the right panel to remove the dependence on galaxy distance and to reduce the effect of stellar mass. The Spearman rank correlation coefficient and the $p$-value are listed in the top left of each panel. } 
\vspace{0.5cm}
\label{fig:sm_and_SFR}
\end{figure*}

\subsection{Impact of Stellar Mass}\label{sec:sm}

In star-forming galaxies, stellar mass is correlated with the star formation activity, gas fraction, molecular-to-atomic gas ratio, and metallicity, and plays a crucial role in setting the gravitational potential of galaxies \citep{2009_Blanton_ARA&A_159B,2021_Leroy_ApJS}. 

For our sample galaxies, the stellar mass ranges from 9.26 $\lesssim$ log $M_{\rm \star}$ [M$_{\odot}$] $\lesssim$ 11.03, with specific star formation rate SFR/$M_{\rm \star}$ $\textgreater$ 10$^{-11}$ yr$^{-1}$. This range selects targets near the $z$ = 0 star-forming main sequence and excludes passive, non-star-forming galaxies, which are less likely to host massive cold gas reservoirs. 

In early-type galaxies, the number and X-ray emission of low-mass X-ray binaries (LMXBs) correlates strongly with stellar mass \citep{2004_Gilfanov_MNRAS_146G,2011_Boroson_ApJ_12B}. Furthermore, recent studies have identified an $L_{\rm X}$$-$SFR$-$$M_{\rm \star}$ scaling relation, which accounts for the contribution of LMXBs and high-mass X-ray binaries, derived from samples of both local and high-redshift galaxies \citep{2010_Lehmer_ApJ_559L,2016_Lehmer_ApJ_7L,2020_Kouroumpatzakis_mnras}. 

All these results indicate that X-ray emission is associated with stellar mass. Therefore, we investigate the correlation between the diffuse X-ray emission of hot gas and stellar mass for the centers of the sample galaxies, as shown in Figure~\ref{fig:sm_and_SFR}. We find that {\lx} and $M_{\rm \star}$ has a relatively strong correlation with $r_{\rm s}$ = 0.67, but it is significantly lower than that of the {\lx}$-$SFR relation (see Table~\ref{tab:table2}). The Bayesian method gives the best-fit relation 
\begin{equation}\label{eq11}
{\rm log}\,(\frac{L_{\rm 0.5-2\,keV}^{\rm gas}} {\rm erg\ s^{-1}}) = 1.51(\pm0.42) {\rm log}\,\frac{M_{\rm \star}} {M_\odot} + 25.13(\pm4.00).  
\end{equation}
The distribution of data points in this relation shows the largest dispersion, with an $rms$ scatter of $\sigma$ = 0.68. 

By combining Equation~\ref{eq8}, the {\lx}$-$$M_{\rm \star}$ relation of the center can be used to obtain the star formation main sequence (SFMS), a well-established relation between SFR and stellar mass. The SFMS derived from our scaling relations is SFR $\thicksim$ $M_{\rm \star}^{1.1}$, which is consistent with previous studies \citep[e.g.,][]{2007_Daddi_ApJ_156D,2016_Saintonge_MNRAS_1749S,2019_Lin_ApJ_33L,2021_Ellison_MNRAS_4777E,2021_Pessa_A&A_134P}.

For the center, we further compare the specific SFR (sSFR) with the diffuse X-ray luminosity normalized by stellar mass in Figure~\ref{fig:sm_and_SFR} (right panel). The best-fit relation between these two parameters is 
\begin{equation}\label{eq12}
{\rm log}\,(\frac{L_{\rm 0.5-2\,keV}^{\rm gas}} {M_{\rm \star}}) = 1.01(\pm0.21) {\rm log}\,\frac{\rm SFR} {M_{\rm \star}} + 40.10(\pm2.06).  
\end{equation}
Compared to the scattered {\lx}$-$$M_{\rm \star}$ relation, the {\lx}/$M_{\rm \star}$$-$sSFR relation is much more compact on the logarithmic scale, with a higher Spearman correlation coefficient of $r_{\rm s}$ = 0.78 and a moderate $rms$ scatter of $\sigma$ = 0.45. The correlation between {\lx}/$M_{\rm \star}$ and sSFR is independent of distance and is less influenced by the intrinsic scaling relation of SFMS. This indicates that physically, the relation of {\lx} with SFR is more fundamental than the {\lx}$-$$M_{\rm \star}$ relation, in agreement with the study of galactic coronae in nearby edge-on galaxies reported by \citet{2013_LiJiangTao_MNRAS_2}. However, it is worth noting that the {\lx}/$M_{\rm \star}$$-$sSFR relation shows larger uncertainties in both the slope and intercept compared to the {\lx}$-$SFR relation, despite having a nominal slope of unity. These uncertainties, particularly in the intercept, may introduce significant systematic error. Therefore, this normalized relation should be treated with caution.

\subsection{Role of Molecular Gas and Baryonic Mass}\label{sec:2mass}

In Section~\ref{sec:sm}, we have shown that diffuse X-ray luminosity in the center is more closely correlated with SFR compared to the stellar mass. This can be attributed to the fact that the hot gas is the product of star formation and subsequent stellar evolution \citep{2010_Kuntz_ApJS_46K}. In the current model of galaxy evolution, star formation occurs within cold, dense molecular gas clouds and is regulated by complex physical processes, such as feedback from young stars and supernovae from the deaths of the massive stars, as well as magnetic fields and hydrostatic pressure exerted by baryonic mass \citep{2004_Gao&Solomon_ApJ_a,2004_Gao&Solomon_ApJS_b,2012_Kennicutt_ARA&A_531K,2019_Krumholz_ARA&A_227K,2020_Chevance_SSRv_50C,2023_Sunjiayi_ApJL_19S}. In this context, hot and cold gas are connected through star formation, with the latter traced by molecular emission lines. 

\begin{figure*}
\centering 
\includegraphics[scale=0.99]{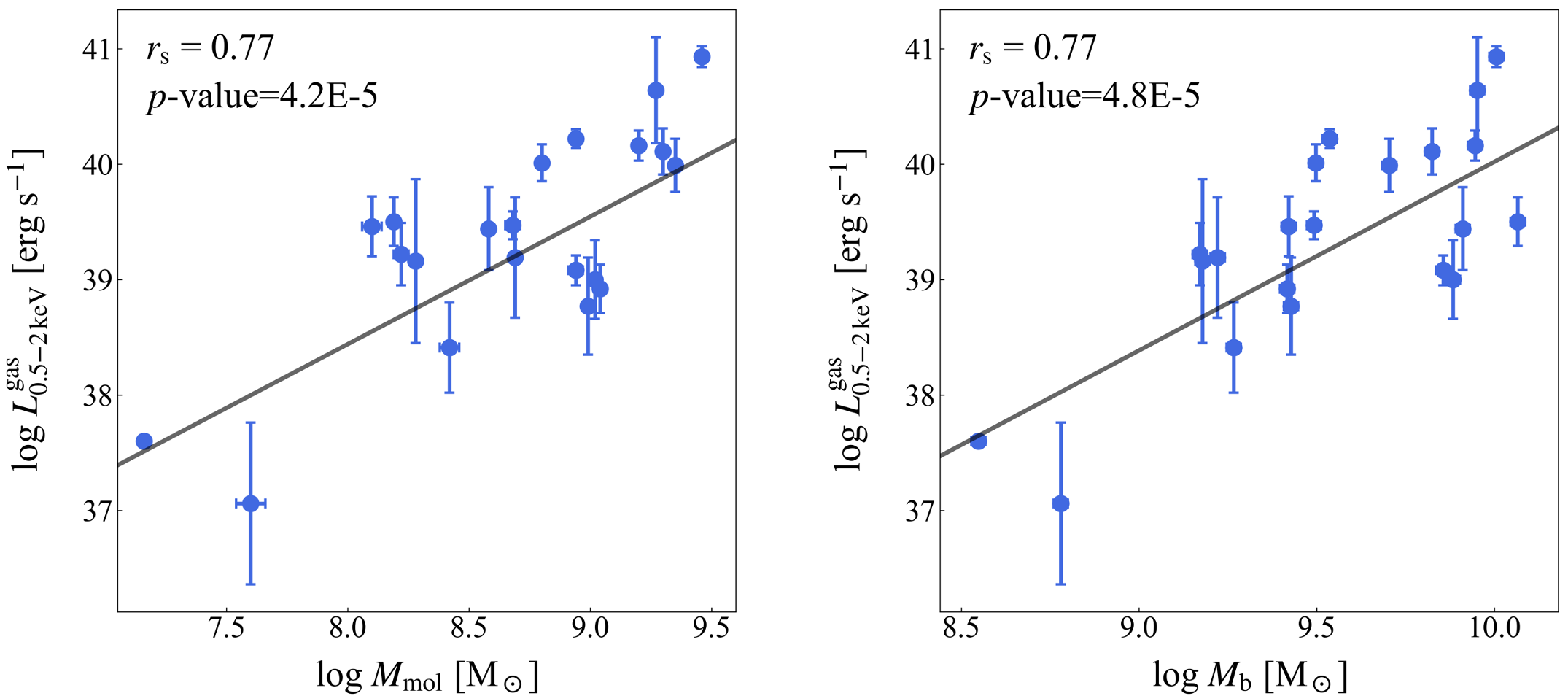}
\vspace{-0.2cm}
\caption{{\lx} as a function of $M_{\rm mol}$ and $M_{\rm b}$, defined as $M_{\rm \star}$ {+} $M_{\rm mol}$, for the centers of the 21 sample galaxies (excluding NGC 1313 and NGC 6744). The central molecular gas mass of these two galaxies can not be obtained due to the extremely faint CO emission of NGC 1313 and the incomplete coverage pattern of NGC 6744, where the bulge has not been observed. } 
\vspace{0.6cm}
\label{fig:mol_and_baryonic}
\end{figure*}

The left panel of Figure~\ref{fig:mol_and_baryonic} shows the correlation between the diffuse X-ray luminosity and the molecular gas mass in the centers of our sample galaxies. We use the CO(2-1) emission line to estimate the molecular gas mass in 21 out of our 23 targets (see Section~\ref{sec:alma}). Due to the faint CO emission of NGC 1313 and the unusual coverage pattern of NGC 6744, where the bulge has not been observed, we can not obtain the molecular gas mass of the centers of these two galaxies. The best-fit result for the {\lx}$-$$M_{\rm mol}$ relation is 
\begin{equation}\label{eq13}
{\rm log}\,(\frac{L_{\rm 0.5-2\,keV}^{\rm gas}} {\rm erg\ s^{-1}}) = 1.13(\pm0.25) {\rm log}\,\frac{M_{\rm mol}} {M_\odot} + 29.57(\pm2.16).  
\end{equation}
The Spearman rank correlation coefficient of this relation is $r_{\rm s}$ = 0.77, close to that of the {\lx}$-$SFR relation, but the $rms$ scatter is $\sigma$ = 0.58. The relatively large scatter in the {\lx}$-$$M_{\rm mol}$ relation reflects the complex physical and chemical processes within the molecular clouds. However, it remains unclear how these processes affect the formation of stars and depend on the galactic environment.

Recent studies of the Milky Way \citep{2010_Lada_ApJ_687L,2012_Lada_ApJ_190L,2014_Evans_ApJ_114E,2016_Vutisalchavakul_ApJ_73V} and external galaxies \citep{2004_Gao&Solomon_ApJ_a,2019_Jimenez_empire} suggest that stars preferentially form in the densest regions of molecular clouds. \citet{2013_Lada_ApJ_133L} and \citet{2014_Evans_ApJ_114E} argue that the overall star formation of a galaxy would be regulated by the amount of dense gas available above a critical density threshold. Another turbulence-regulated model considers the global properties of molecular clouds as the key factors determining their density distribution \citep[e.g.,][]{2007_Krumholz_ApJ_289K,2012_Federrath_ApJ_156F}. In this scenario, star formation and its efficiency depend on cloud properties. On the other hand, the spatial distribution of soft diffuse X-ray emission is found to be correlated with the sites of recent star formation in spiral arms \citep{2004_Tyler_ApJ}. This implies that the thermal emission of hot gas could be regulated by these cloud parameters.

The mid-plane pressure of the interstellar medium is an important role influencing the properties of molecular clouds. \citet{2019_Schruba_ApJ_2S} and \citet{2020_Sunjiayi_ApJ_148S} found that the average internal pressure of molecular clouds tends to balance the weight of the clouds themselves and the external pressure from the interstellar medium. In hydrostatic equilibrium, the mid-plane pressure in a galactic disk will adjust to support the gravity of the galaxy. Therefore, in this work, we define the baryonic mass as a proxy for the mid-plane pressure.

\citet{2021_Barrera-Ballesteros_ApJ_131B} computed the baryonic mass surface density as $\Sigma_{\rm b} = \Sigma_{\star} + \Sigma_{\rm mol}$ for $\thicksim$2600 galaxies and found that $\Sigma_{\rm b}$ tightly correlates with SFR, where $\Sigma_{\star}$ and $\Sigma_{\rm mol}$ are the stellar surface density and the molecular gas mass surface density, respectively. Following their work, we explore the correlation between the diffuse X-ray emission and the baryonic mass $M_{\rm b}$ at the centers of our sample galaxies (except for NGC 1313 and NGC 6744). The right panel of Figure~\ref{fig:mol_and_baryonic} shows the {\lx}$-$$M_{\rm b}$ relation, and the Bayesian method gives
\begin{equation}\label{eq14}
{\rm log}\,(\frac{L_{\rm 0.5-2\,keV}^{\rm gas}} {\rm erg\ s^{-1}}) = 1.65(\pm0.38) {\rm log}\,\frac{M_{\rm b}} {M_\odot} + 23.67(\pm3.65).  
\end{equation}
The $rms$ scatter of the {\lx}$-$$M_{\rm b}$ relation is $\sigma$ = 0.58, which is higher than that measured for the {\lx}$-$SFR relation. This suggests that star formation rate is a better predictor of {\lx} than baryonic mass or mid-plane pressure. We argue that the {\lx}$-$$M_{\rm mol}$ and {\lx}$-$$M_{\rm b}$ relations could originate from the underlying relation between {\lx} and SFR.

\subsection{Contamination of Unresolved Sources}\label{sec:Contamination}

For extragalactic galaxies, several intrinsically faint X-ray sources are difficult to detect with current instruments. These sources include unresolved LMXBs, coronally active binaries (ABs), cataclysmic variables (CVs), and young faint objects, including protostars, young stellar objects, and young stars. Typically, a single type of faint compact source contributes no more than 10\% of the soft X-ray emission, and most contribute less than 5\% \citep{2010_Kuntz_ApJS_46K,2021_Wang_M83_6155W}. However, the collective contribution of these sources may be non-negligible for regions at $\thicksim$1 kpc scales. 

To determine the total contribution of these faint sources below the detection limits, we employ several complementary methods to quantify their X-ray flux in the 0.5$-$2 keV band. The emission of unresolved LMXBs, along with the collective contribution from ABs and CVs, scales closely with stellar mass. Thus, the $K$-band image, which traces the stellar mass distribution, can be used to estimate the luminosities of these sources in the 0.5$-$2 keV band, according to the prescriptions of \citet{2011_Boroson_ApJ_12B}:
\begin{equation}\label{eq15}
L_{\mathrm{X}}(\mathrm{ABs+CVs})/L_{K}=4.4_{-0.9}^{+1.5} \times 10^{27} \, \mathrm{erg \, s^{-1}} L_{\mathrm{K\odot}}^{\, \, -1}, 
\end{equation}
\begin{equation}\label{eq16}
L_{\mathrm{X}}(\mathrm{LMXBs})/L_{K}=10^{29.0 \, \pm \, 0.176} \, \mathrm{erg \, s^{-1}} L_{\mathrm{K\odot}}^{\, \, -1}, 
\end{equation}
where $L_{\mathrm{K\odot}}$ is in solar luminosity. For young faint objects, their X-ray luminosity in the 2$-$10 keV band is found to be correlated with SFR \citep{2011_Bogdan_MNRAS}
\begin{equation}\label{eq17}
    L_{\mathrm{X}}(\mathrm{young})/{\rm SFR}=(1.7 \pm 0.9) \times 10^{38} (\mathrm{erg \, s^{-1}})/({\rm M_{\odot}\ yr^{-1}}). 
\end{equation}
We use this relation to estimate the hard X-ray emission of this component. We then convert the 2$-$10 keV luminosity to the 0.5$-$2 keV band using a conversion factor of $L_{\rm 0.5 - 2\,keV}$/$L_{\rm 2 - 10\,keV}$ = 0.64 from \citet{2025_Kyritsis_A&A_128K}.

\begin{figure}
\includegraphics[scale=0.523]{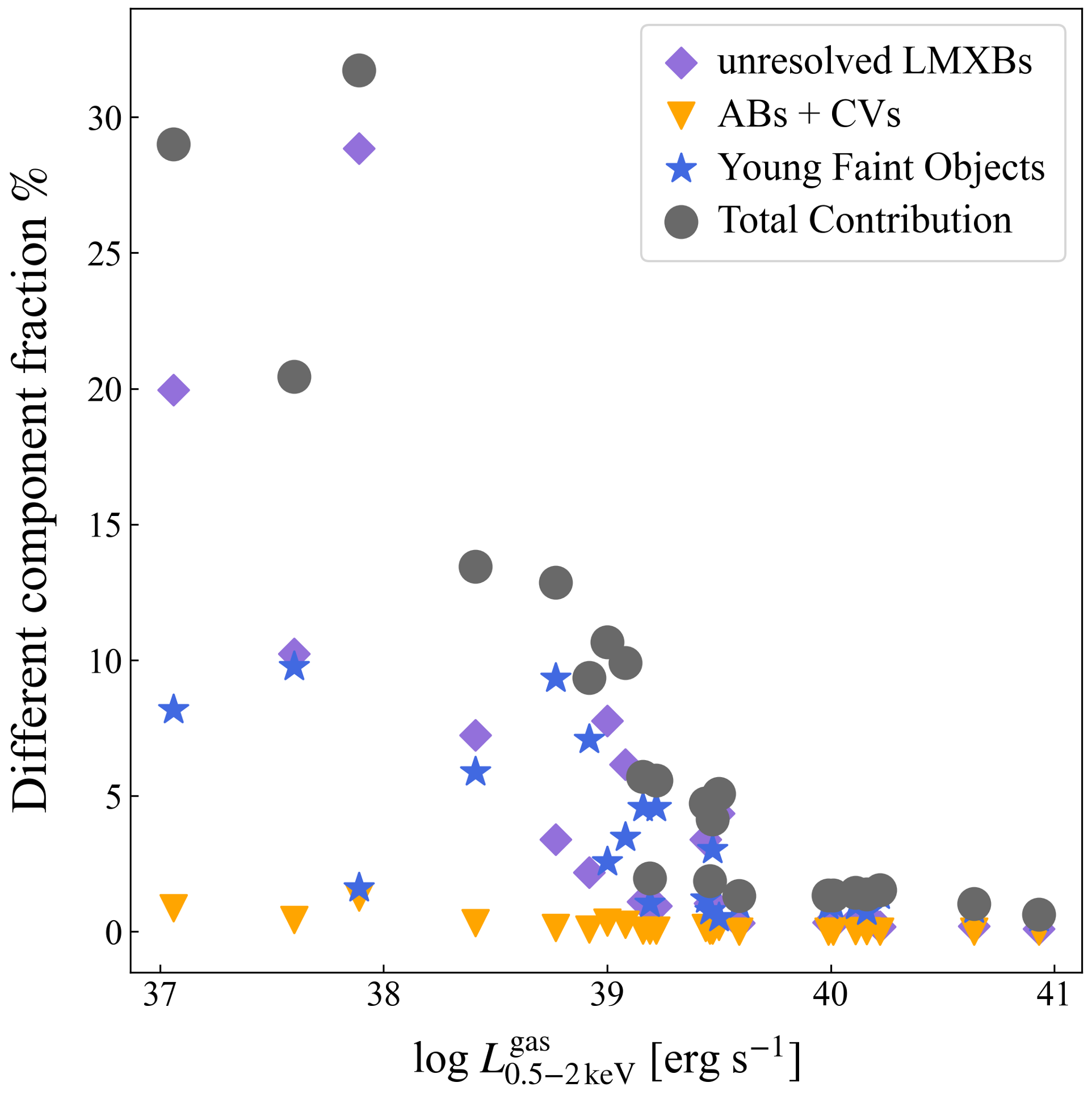}
\vspace{-0.5cm}
\caption{The contribution fraction of unresolved emission from faint compact sources relative to the soft X-ray luminosity in the 0.5$-$2 keV band. The contributions of unresolved LMXBs (\emph{purple diamonds}) and young faint objects (\emph{blue stars}), as well as the total contribution (\emph{grey circles}) of all faint sources, anticorrelate with {\lx}, while the contribution of ABs and CVs (\emph{orange triangles}) remains stable and negligible.  } 
\vspace{0.2cm}
\label{fig:fraction}
\end{figure}

Figure~\ref{fig:fraction} shows the contribution fraction of each type of faint compact source and their total emission to the soft X-ray luminosity. The fraction of ABs and CVs is the smallest compared to other types of sources for all galaxies, contributing only $\thicksim$0.1$-$2\% of the X-ray emission. For most galaxies, the fraction of young faint objects is comparable to that of unresolved LMXBs. Moreover, we find that the contribution fractions of young faint objects and unresolved LMXBs, as well as the total contribution, anticorrelate with the soft X-ray luminosity. This result implies that the emission from these undetected faint sources is relatively stable in our sample galaxies. The average contribution of all galaxies is $\thicksim$8\%, with only three targets exceeding 20\% but remaining below 35\% due to their low X-ray luminosities, consistent with the results estimated from previous studies \citep[e.g.,][]{2010_Kuntz_ApJS_46K,2012_Mineo_hotgas,2021_Wang_M83_6155W}. 

Given the minor contributions of unresolved LMXBs, young faint objects, and ABs and CVs, we conclude that the source-subtracted X-ray emission in the 0.5$-$2 keV band is a reasonable approximation to the emission of hot gas in the center. 

\subsection{X-ray Radiation Efficiency and Stellar Feedback}\label{sec:Efficiency}

The X-ray radiation efficiency, {\lx}/$\dot{E}_\mathrm{SN}$, of star-forming galaxies is an important tool for studying stellar feedback, which includes processes such as SN explosions and stellar winds. Recent studies have found that core-collapse SNe dominate the stellar mechanical energy input in the central regions of spiral galaxies \citep[see][]{2012_Mineo_hotgas,2025_Zhangcy_ApJ_15Z}. Thus, assuming an explosion rate of core-collapse SNe of $\thicksim$ 1 SN per 100 yr per 1 $M_{\odot}$/yr \citep{2012_Botticella_A&A}, and a mechanical energy release of $E_\mathrm{SN}$ = 10$^{51}$ erg per explosion \citep{2005_Grimes_ApJ_187G,2008_Bogd_MNRAS}, we can obtain a SN mechanical energy injection rate of $\dot{E}_\mathrm{SN}$ $\approx$ 3.2 $\cdot$ 10$^{41}$ $\times$ SFR erg/s. By comparing the hot gas luminosity with the derived $\dot{E}_\mathrm{SN}$, we then estimate the X-ray radiation efficiencies in the central regions to range from $\thicksim$1\% to $\thicksim$10\%, consistent with previous results \citep[e.g.,][]{2012_Mineo_hotgas,2013_LiJiangTao_MNRAS_2,2025_Zhangcy_ApJ_15Z,2025_Luan_apj}.

The very low radiation efficiencies suggest that a significant fraction of the mechanical energy from stellar feedback is dissipated or escapes without being radiated in the soft X-ray band. Based on a lognormal temperature distribution model, \citet{2025_Luan_apj} found that the majority of the thermal emission in M51 may occur in the 0.01$-$0.3 keV energy band, predominantly originating from plasma at temperatures below 0.1 keV. In their model, the galactic corona contributes $\thicksim$74\% of the total radiative energy emitted by the hot plasma over the 0.01$-$100 keV band, with the remaining quarter coming from the hot gas in the disk. Furthermore, the total radiative energy of the hot plasma is about 15 times greater than the energy emitted in the 0.5$-$2 keV band. This indicates that the energy of stellar feedback is mainly released by the cooling of the hot plasma in the galaxy.

The cooling of hot interstellar gas is also observed in cooling-flow clusters, where multiphase filaments extending from the central galaxy may arise from the condensation of hot gas \citep[e.g.,][]{2001_Conselice_AJ_2281C, 2003_Salom_A&A_657S, 2012_McDonald_Natur_349M, 2019_Vantyghem_ApJ_57V}. \citet{2025_Olivares_NatAs_449O} quantified a tight linear correlation between diffuse X-ray and H${\rm \alpha}$ surface brightness in the filaments of seven strong cooling-flow clusters. This discovery provides evidence for a shared excitation mechanism between hot and warm gas in filaments, where multiphase condensation appears to drive their co-evolution.

\section{Summary}\label{sec:summary}

We have used high spatial resolution data from $Chandra$, WISE, and GALEX to derive SFR and diffuse X-ray luminosity of hot gas in a sample of 23 nearby star-forming galaxies and to study the {\lx}$-$SFR scaling relation in the central regions of these galaxies across multiple spatial scales (with radii of 0.5, 1, 1.5, 2, 2.5, and 3 kpc). We investigated the relationship between hot gas and molecular gas based on ALMA data from the PHANGS survey. Additionally, we explored the {\lx}$-$$M_{\rm stellar+mol.gas}$ correlation in the galactic center to determine the effect of mid-plane hydrostatic pressure on the central hot gas emission. We applied different methods to estimate the contribution of unresolved emission from faint compact sources, including unresolved LMXBs, young faint objects, as well as ABs and CVs, to the soft X-ray luminosity in the 0.5$-$2 keV band. Our main results are summarized as follows: 

\begin{enumerate}
\item We used a two-component $\beta$-model to fit the surface brightness profiles of point source-excluded X-ray maps of our 23 sample galaxies. All galaxies show a break point, defined as the intersection point of the two $\beta$-models, in the surface brightness profile at a distance of 0.3$-$2.2 kpc. The surface brightness declines sharply within the break radius but varies gradually beyond it. This implies that the hot gas has different origins or heating mechanisms in the inner region and the galactic disk.

\item The slope of the relation between SFR and diffuse X-ray luminosity vary significantly in different regions of the sample galaxies. A super-linear {\lx}$-$SFR relation is observed in the centers, while a sub-linear relation is found in the disk. For entire galaxies, the global linear correlation is consistent with previous findings.

\item We found that the {\lx}$-$SFR relation depends on spatial scale in the galactic central region. Its slope anticorrelates with spatial scale up to 2.5 kpc, where it drops to the global level. This suggests that the characteristics of central hot gas emission are entirely erased on a scale of $\thicksim$2.5 kpc in radius.

\item We searched for the correlation between central diffuse X-ray luminosity and stellar mass. We found a relatively strong relation between these two physical quantities but with the largest scatter compared to other relations.

\item The molecular gas mass and baryonic mass present good correlations with the diffuse X-ray luminosity in the center. The latter implies that the mid-plane pressure has an effect on the emission of hot gas. However, the scatters of the {\lx}$-$$M_{\rm mol}$ and {\lx}$-$$M_{\rm b}$ relations are larger than that of the {\lx}$-$SFR relation. We conclude that the star formation rate is a better predictor for diffuse X-ray emission than stellar mass, molecular gas, and baryonic mass.

\item The contribution of faint compact sources to the soft X-ray emission in the 0.5$-$2 keV band is $\thicksim$8\% on average and can be safely ignored. Among these sources, ABs and CVs contribute the least compared to unresolved LMXBs and young faint objects, with a fraction of only $\thicksim$0.1-2\%. 

\end{enumerate}

Studying diffuse X-ray emission of hot gas at high spatial resolution is a powerful bridge between extragalactic observations and measurements within our own Galaxy. In the future, with significant observational efforts devoted to probing more extreme environmental conditions (e.g., very high or very low surface densities), we could build a large, homogeneous dataset to provide deeper insights into the growth and evolution of galaxies in the local universe.

We thank the anonymous referee for providing constructive comments that improved our work. C.Y.Z. thanks Dr. Xue-Jian Jiang and Shui-Nai Zhang for their helpful suggestions. J.W. acknowledges the National Key R$\&$D Program of China (grant No. 2023YFA1607904) and the National Natural Science Foundation of China (NSFC) grants 12033004, 12221003, and 12333002 and the science research grant from CMS-CSST-2025-A10 and CMS-CSST-2025-A07. This research made use of Chandra archival data and software provided by the Chandra X-ray Center (CXC) in the application package CIAO. This research has made use of SAOImage DS9, developed by Smithsonian Astrophysical Observatory. CARTA is the Cube Analysis and Rendering Tool for Astronomy, a new image visualization and analysis tool designed for the ALMA, the VLA, and the SKA pathfinders.

This work makes use of data products from the Wide-field Infrared Survey Explorer (WISE), which is a joint project of the University of California, Los Angeles, and the Jet Propulsion Laboratory/California Institute of Technology, funded by NASA. This work is based in part on observations made with the Galaxy Evolution Explorer (GALEX). GALEX is a NASA Small Explorer, whose mission was developed in cooperation with the Centre National d'Etudes Spatiales (CNES) of France and the Korean Ministry of Science and Technology. This work makes use of data products from the PHANGS$-$ALMA CO(2$-$1) survey. ALMA is a partnership of ESO (representing its member states), NSF (USA) and NINS (Japan), together with NRC (Canada), MOST and ASIAA (Taiwan), and KASI (Republic of Korea), in cooperation with the Republic of Chile. The Joint ALMA Observatory is operated by ESO, AUI/NRAO and NAOJ. The National Radio Astronomy Observatory is a facility of the National Science Foundation operated under cooperative agreement by Associated Universities, Inc.


\bibliography{bibliography}{}
\bibliographystyle{aasjournal}

\end{document}